\documentclass[onecolumn,showpacs,preprintnumbers,amsmath,amssymb]{revtex4}
\usepackage{graphicx}% Include figure files
\usepackage{dcolumn}% Align table columns on decimal point
\usepackage{bm}% bold math

\def\In{\displaystyle \int}
\def\sqr#1#2{{\vcenter{\hrule height.#2pt\hbox{\vrule width.#2pt
height#1pt \kern#1pt \vrule width.#2pt}\hrule height.#2pt}}}
\def\square{\mathchoice\sqr64\sqr64\sqr{4.2}3\sqr{3.0}3}
\begin{document}

%\preprint{VA-BAO-11-2013}

\title{Is there an inertia due to the supersymmetry}

\author{G Ter-Kazarian}

 \email{gago_50@yahoo.com}
\affiliation{
Byurakan Astrophysical Observatory\\
Byurakan 378433, Aragatsotn District, Armenia}

%\date{\today}% It is always \today, today,
            %  but any date may be explicitly specified

\begin{abstract}
We derive a standard Lorentz code (SLC) of motion  by exploring
rigid double transformations of, so-called, {\it master
space-induced} supersymmetry (MS-SUSY), subject to certain rules.
The renormalizable and actually finite flat-space field theories
with $N_{max}=4$ supersymmetries in four dimensions, if only such
symmetries are fundamental to nature, yield the possible {\it
extension of Lorentz code} (ELC), at which the SLC violating new
physics appears.  In the framework of local MS-SUSY, we address the
inertial effects. We argue that a space-time deformation of MS is
the origin of inertia effects that can be observed by us. We go
beyond the hypothesis of locality. This allows to improve the
relevant geometrical structures referred to the noninertial frame in
Minkowski space for an arbitrary velocities and characteristic
acceleration lengths. This framework furnishes justification for the
introduction of the {\it weak} principle of equivalence, i.e., the
{\it universality of free fall}. The implications of the inertia
effects in the more general post-Riemannian geometry are briefly
discussed.
\end{abstract}


%\keywords{Space-Time Symmetries, OPERA Superluminal neutrinos, Inertia, Spacetime Deformation,
%Principle of Equivalence, Noninertial Frames, Post-Riemannian
%Geometry}

\maketitle
\section{Introduction}
The principle of inertia, whose origin can be traced back to the
works developed by Galileo~\cite{Drake} and Newton~\cite{Newt}, is
one of the fundamental principles of the classical mechanics. This
governs  the {\it uniform motion}  of a body and describes how it is
affected by applied forces.  The universality of gravitation and
inertia attribute to the geometry but as having a different natures.
However, despite the advocated success of general relativity (GR),
the problem of inertia stood open and that this is still an unknown
exciting problem to be challenged. The inertia effects cannot be in
full generality identified with gravity within GR as it was proposed
by Einstein in 1918 \cite{Nort}, because there are many experimental
controversies to question the validity of such a description, for
details see e.g.~\cite{gago1} and references therein. The model
discussed in the latter illustrates the problems of inertia effects,
but it also hints at a possible solution. We will not be concerned
with the actual details of this model here, but only use it as a
backdrop to explore first the SLC in a new perspective of rigid
double transformations of, so-called, {\it master space-induced}
supersymmetry (MS-SUSY), subject to certain rules. The theories with
extended $N_{max}=4$ supersymmetries, namely $N=4$ super-Yang-Mills
theories, if only such symmetries are fundamental to nature, lead to
the model of ELC in case of the apparent violations of SLC, the
possible manifestations of which arise in a similar way in all
particle sectors. We show that in the ELC-framework the propagation
of the superluminal particle could be consistent with causality, and
give a justification of forbiddance of Vavilov-Cherenkov
radiation/or analog processes in vacuum. However, we must be careful
about the physical relevance of the standard theory of extended
supersymmetry which does not allow for chiral fermions, and that its
spectrum in no way resembles that of the observed in
nature~\cite{Soh}. Consequently, in the framework of local MS-SUSY,
we address the {\it accelerated motion}, while, unlike gravitation,
a curvature of space-time now arises entirely due to the inertial
properties of the Lorentz-rotated frame of interest, i.e. a {\it
fictitious gravitation} which can be globally removed by appropriate
coordinate transformations. The only source of graviton and
gravitino, therefore, is the acceleration of a particle. This paper
is organized as follows. In the next section, we explain our idea of
what is the MS. In section 3, we give a hard look at MS. The MS-SUSY
is dealt with in section 4. In section 5, the complementary approach
is developed where the SLC and ELC are derived. More about the
accelerated motion is said in section 6, this time in the presence
of the local MS-SUSY. In section 7 we briefly discuss the inertia
effects. We go then beyond the hypothesis of locality in subsections
A-C.  We compute the improved metric and other relevant geometrical
structures in noninertial system of arbitrary accelerating and
rotating observer in Minkowski space-time. The case of semi-Riemann
background space $V_{4}$ is studied in subsection D, whereas we give
justification for the introduction of the {\it weak} principle of
equivalence (WPE) on the theoretical basis, which establishes the
independence of free-fall trajectories of the internal composition
and structure of bodies. The implications of the inertial effects in
the more general post-Riemannian geometry are briefly discussed in
subsection E. The concluding remarks are presented in section 8. We
will be brief and often ruthlessly suppress the indices without
notice. Unless otherwise stated we take natural units, $h=c=1$.

\section{Preliminaries: \,\,A first glance at MS}
With regard now to our original question as to the  understanding of
the physical processes that underly the {\it motion}, we tackle the
problem in the framework of quantum field theory. Let us consider
functional integrals for a quantum-mechanical system with one degree
of freedom. Denote by $x(t)$ the position operator in the Heisenberg
picture, and by $|x,t>$ its eigenstates. The probability amplitude
that a particle which was at $x$ at time $t$ will be at point $x'$
at time $t'$, also called the Schwinger transformation function for
these points, is $F(x' t'; xt)=<x't'\,|xt>$.  For a particle moving
through the two infinitesimally closed points of original space,
this  in somehow or other implies the elementary act consisting of
the {\it annihilation} of a particle at the point $x$ and time $t$
and, subsequently, its {\it creation} at the point $x'$ and time
$t'$. The particle can move with different velocities which
indicates to existence of the intermediate, so-called, {\it motion}
state. Then the {\it annihilation} of a particle at point $x$ and
time $t$ can intuitively be understood as the transition from the
initial state $|x,t>$ to the intermediate {\it motion} state,
$|\underline{x},\underline{t}>$, yet unknown, where
$\underline{x}(\underline{t})$ represent atomic element of idealized
{\it motion} point event. Meanwhile, the {\it creation} of a
particle at infinitesimally closed final point $x'$ and time $t'$
means the subsequent transition from the intermediate {\it motion}
state, $|\underline{x},\underline{t}>$, to the final state,
$|x',t'>$.  So, the Schwinger transformation function for two
infinitesimally closed points is written in terms of annihilation
and creation processes of a particle as
\begin{equation}
\begin{array}{l}
F(x' t'; xt)=\In\,d\underline{x}<x'
t'\,|\underline{x}\,\underline{t}><\underline{x}\,\underline{t} \,|x
t>.
\end{array}
\label{TR1}
\end{equation}
It should be emphasized that since we do not understand the
phenomenon of {\it motion}, then here it must suffice to expect that
the state functions $|x,t>$ and $|\underline{x},\underline{t}>$ are
quite different. Therefore, the intermediate {\it motion} state,
$|\underline{x},\underline{t}>$,  can be defined on say {\it motion}
space, $\underline{M}$, the points $\underline{x}(\underline{t})$ of
which are all the {\it motion} atomic elements,
$(\underline{x}(\underline{t})\in \underline{M})$. To express
Schwinger transformation function, $F$, as a path integral, we
divide the finite time interval into $n+1$ intervals: $ t=t_{0},
t_{1},\dots,t_{n+1}=t';\quad t_{k}=t_{0}+k\varepsilon,$ where
$\varepsilon$ can be made arbitrarily small by increasing $n$. In
the limit $n\rightarrow\infty$, by virtue of ~(\ref{TR1}), $F$
becomes an operational definition of the path integral. Hence, in
general, in addition to background 4D Minkowski space $M_{4}$,  also
a background {\it motion} space $\underline{M}$, or say {\it master
space}, MS \,$(\equiv \underline{M})$ is required. So, we now
conceive of the two different spaces $M_{4}$ and MS, where the
geometry of MS is a new physical entity, with degrees of freedom and
a dynamics of its own. The above example ~(\ref{TR1}) imposes a
constraint upon MS that it was embedded in $M_{4}$ as an
indispensable individual companion to the particle, without relation
to the other matter. In going into practical details, we further
adopt the model discussed recently in reference~\cite{gago1}, which
illustrates the problems of inertia effects, but it also hints at a
possible solution. In accord, MS is not measurable directly, but it
was argued that a deformation~(also see~\cite{gago2}) of MS is the
origin of inertia effects that can be observed by us. In general
case of 3D motion in $M_{4}$, following~\cite{gago1}, a flat MS is
the 2D Minkowski space $\underline{M}_{\,2}$ (see next section). In
deriving the final step, we should compare and contrast the particle
states of quantum fields defined on the background spaces $M_{4}$
and $\underline{M}_{\,2}$, forming a basis in the Hilbert space. It
is quite clear that the following properties, being the essence of
the chain of transformations~(\ref{TR1}) for the finite time
interval, hold:

1. There should be a particular way of going from each point
$x_{i-1}(t_{i-1})\in M_{4}$ to the intermediate {\it motion} point
$\underline{x}_{\,i-1}(\underline{t}_{\,i-1})\in
\underline{M}_{\,2}$ and back $x_{i}(t_{i})\in M_{4}$, such that the
net result of each atomic double transformations is as if we had
operated with a space-time {\it translation} on the original space
$M_{4}$.  So, the symmetry we are looking for must mix the particle
quantum states during the motion in order to reproduce the central
relationship between the two successive transformations of this
symmetry and the generators of space-time translations. Namely, the
subsequent operation of two finite transformations will induce a
translation in space and time of the states on which they operate.

2. These successive transformations induce in $M_{4}$ the
inhomogeneous Lorentz group, or Poincar\'{e} group, and that a
unitary linear transformation   $|x,t>\rightarrow U(\Lambda,
a)|x,t>$  on vectors in the physical Hilbert space.

Thus, the underlying algebraic structure of this symmetry generators
closes with the algebra of {\it translations} on the original space
$M_{4}$ in a way that it can then be summarized as a non-trivial
extension of the Poincar\'{e} group algebra, including the
generators of translations. The only symmetry possessing such
properties is the supersymmetry (SUSY), see
e.g.~\cite{DHM}-\cite{DM}, which is accepted as a legitimate feature
of nature, although it has never been experimentally observed.
Certainly we now need to modify the standard theory to have MS-SUSY,
involving  a superspace which is an enlargement of a direct sum of
background spaces $M_{4}\oplus\underline{M}_{\,2}$ by the inclusion
of additional fermion coordinates. Thereby an attempt will be made
to treat the {\it uniform motion} of a particle as a complex process
of the global (or rigid) MS-SUSY double transformations. Namely a
particle undergoes to {\em an infinite number of successive
transitions from $M_{4}$ to $\underline{M}_{\,2}$ and back going
permanently through fermion-boson transformations}, which can be
interpreted as its {\it creation} and {\it annihilation} processes
occurring  in $M_{4}$ or $\underline{M}_{\,2}$. We derive the {\it
Lorentz code} of motion in terms of spinors referred to MS. This
allows to introduce the physical finite {\it time interval} between
two events, as integer number of the {\it duration time} of atomic
double transition of a particle from $M_{4}$ and back. While all the
particles are living on $M_{4}$, their superpartners can be viewed
as living on $\underline{M}_{\,2}$.

\section{A hard look at MS}
Following ~\cite{gago1}, we assume that a flat MS is the 2D
Minkowski space:
\begin{equation}
\begin{array}{l}
\underline{M}_{\,2}=R_{(+)}^{1}\oplus R_{(-)}^{1}.
\end{array}
\label{S1}
\end{equation}
The ingredient 1D-space $ R_{\underline{m}}^{1}$ is spanned by the
coordinates $\eta^{\underline{m}}$. The following notational
conventions are used throughout this paper:  \, all magnitudes
related to the space $\underline{M}_{\,2}$  will be underlined. In
particular, the underlined lower case Latin letters $\underline{m},
\underline{n},...=(\pm)$ denote the world indices related to
$\underline{M}_{\,2}$. The metric in $\underline{M}_{\,2}$ is
\begin{equation}
\begin{array}{l}
\underline{g}=\underline{g}(\underline{e}_{\,\underline{m}},\,\underline{e}_{\,\underline{n}})\,
\underline{\vartheta}^{\,\underline{m}} \otimes
\underline{\vartheta}^{\,\underline{n}},
\end{array}
\label{S2}
\end{equation}
where
$\underline{\vartheta}^{\,\underline{m}}=d\eta^{\,\underline{m}}$ is
the infinitesimal displacement. The basis
$\underline{e}_{\,\underline{m}}$ at the point of interest in
$\underline{M}_{\,2}$ is consisted of the two real {\em null
vectors}:
\begin{equation}
\begin{array}{l}
\underline{g}(\underline{e}_{\,\underline{m}},\,\underline{e}_{\,\underline{n}})
\equiv<\underline{e}_{\,\underline{m}},\,\underline{e}_{\,\underline{n}}>={}^{*}o_{\,\underline{m}
\,\underline{n}}, \quad
({}^{*}o_{\,\underline{m}\,\underline{n}})=\left(
                                                       \begin{array}{cc}
                                                         0 & 1 \\
                                                         1 & 0 \\
                                                       \end{array}
                                                     \right)
.
\end{array}
\label{S3}
\end{equation}
The norm, $i\overline{d}\equiv d\hat{\eta}$, given in the
basis~(\ref{S3}) reads
$i\underline{d}=\underline{e}\underline{\vartheta}=\underline{e}_{\,\underline{m}}\otimes\underline{\vartheta}^{\,\underline{m}}$,
where $i\underline{d}$ is the tautological tensor field of type
(1,1),  $\underline{e}$ is a shorthand for the collection of the
2-tuplet $(\underline{e}_{(+)},\,\underline{e}_{(-)})$, and $
\underline{\vartheta}=\left(
                    \begin{array}{c}
                      \underline{\vartheta}{}^{(+)} \\
                      \underline{\vartheta}{}^{(-)} \\
                    \end{array}
                  \right).
$ We may equivalently use a temporal $q^{0}\in T^{1}$ and  a spatial
$q^{1}\in R^{1}$ variables $q^{r} (q^{0},\, q^{1}) (r=0,1),$ such
that
\begin{equation}
\begin{array}{l}
\underline{M}_{\,2}=R^{1}\oplus T^{1}.
\end{array}
\label{S4}
\end{equation}
The norm, $i\underline{d}$,  now can be rewritten in terms of
displacement, $dq^{r}$, as
\begin{equation}
\begin{array}{l}
i\underline{d}=d\hat{q}= e_{0}\otimes dq^{0} + e_{1}\otimes dq^{1},
\label{S5}
\end{array}
\end{equation}
where $e_{0}$ and $e_{1}$ are, respectively, the temporal and
spatial basis vectors:
\begin{equation}
\begin{array}{l}
e_{0}=\frac{1}{\sqrt{2}}\left(\underline{e}_{(+)}+\underline{e}_{(-)}\right),\quad
e_{1}=\frac{1}{\sqrt{2}}\left(\underline{e}_{(+)}-\underline{e}_{(-)}\right),
\\
\underline{g}(e_{r},\,e_{s})\equiv<e_{r},\,e_{s}>=o_{r s}, \quad
(o_{r s})=\left(
                                                       \begin{array}{cc}
                                                         1 & 0 \\
                                                         0 & -1 \\
                                                       \end{array}
                                                     \right).
\end{array}
\label{S6}
\end{equation}
The $\underline{M}_{\,2}$-companion  is  smoothly (injective and
continuous) embedded in the $M_{4}$. Suppose the position of the
particle in the background $M_{4}$ space is specified by the
coordinates $x^{m}(s)$ $ (m=0,1,2,3) (x^{0}=t)$ with respect to the
axes of the inertial system $S_{(4)}.$ Then, a smooth map
$f:\phantom{a}\underline{M}_{2}\,\longrightarrow\,M_{4}$ is defined
to be an immersion\,-\,  an embedding which is a function that is a
homeomorphism onto its image:
\begin{equation}
\begin{array}{l}
q^{0}=\frac{1}{\sqrt{2}}\left(\eta^{(+)}+\eta^{(-)}\right)=t,\quad
q^{1}=\frac{1}{\sqrt{2}}\left(\eta^{(+)}-\eta^{(-)}\right)=|\vec{x}|.
\end{array}
\label{S7}
\end{equation}
To motivate why is the MS two dimensional, we note that only two
dimensional constructions of real {\em null vectors}~(\ref{S6}) are
allowed as the basis at given point in MS, which can be embedded in
the (3+1)-dimensional spacetime. This theory is mathematically
somewhat similar to the more recent membrane theory, so the
$\underline{M}_{\,2}$ can be viewed as 2D space living on the 4D
world sheet. Given the inertial frame $S_{(4)}$ in $M_{4}$, we may
define the corresponding inertial frame $S_{(2)}$  used by the
non-accelerated observer for the position $q^{r}$ of a free particle
in flat $\underline{M}_{\,2}$. Thereby the time axes of the two
systems $S_{(2)}$ and $S_{4}$ coincide in direction and that the
time coordinates are taken the same, $q^{0}=t$. For the case at
hand,
\begin{equation}
\begin{array}{l}
v^{(\pm)}=\frac{d\eta^{(\pm)}}{dq^{0}}=\frac{1}{\sqrt{2}}(1\pm
v_{q}), \quad
v_{q}=\frac{dq^{1}}{dq^{0}}=|\vec{v}|=|\frac{d\vec{x}}{dt}|.
\end{array}
\label{S8}
\end{equation}
So the particle may be viewed as moving simultaneously in $M_{4}$
and  $\underline{M}_{\,2}$. Hence, given the inertial frames
$S_{(4)}$, $S'_{(4)}$, $S''_{(4)}$,... in $M_{4}$, in this manner we
may define the corresponding inertial frames $S_{(2)}$, $S'_{(2)}$,
$S''_{(2)}$,... in $M_{2}$. Suppose the elements of the Hilbert
space can be generated by the action of field-valued operators
$\phi(x)\,(\chi(x),\,A(x))$\,\, ($x\in M_{4}$), where $\chi(x)$ is
the Weyl fermion and $A(x)$ is the complex scalar bosonic field
defined on $M_{4}$, and accordingly, of field-valued operators
$\underline{\phi}(\eta)\,(\underline{\chi}(\eta),\,\underline{A}(\eta))$
\,\, ($\eta\in \underline{M}_{2}$), where $\underline{\chi}(\eta)$
is the Weyl fermion and $\underline{A}(\eta)$ is the complex scalar
bosonic field defined on $\underline{M}_{\,2}$, on a translationally
invariant vacuum:
\begin{equation}
\begin{array}{l}
 |x>=\phi(x)|0>,\quad |x_{1}, x_{2}>=\phi(x_{1})\phi(x_{2})|0>
\quad\mbox{referring to}\quad
 M_{4},\\
 |\eta>=\underline{\phi}(\eta)|0>,\quad |\eta_{1},
\eta_{2}>=\underline{\phi}(\eta_{1})\underline{\phi}(\eta_{2})|0>
\,\,\,\quad\mbox{referring to}\quad \underline{M}_{\,2},
\end{array}
\label{S9}
\end{equation}
etc. The displacement of the field takes the form
\begin{equation}
\begin{array}{l}
\phi(x_{1}+x_{2})=e^{ix_{2}^{\,m}P_{m}}\,\phi(x_{1})\,e^{-ix_{2}^{\,m}P_{m}},\quad
\underline{\phi}(\eta_{1}+\eta_{2})=e^{i\eta_{2}^{\,\underline{m}}P_{\underline{m}}}
\,\underline{\phi}(\eta_{1})\,e^{-i\eta_{2}^{\,\underline{m}}P_{\underline{m}}},
\end{array}
\label{S10}
\end{equation}
where $P_{m}=i\partial_{m}$ is the generator of translations on
quantum fields $\phi(x)$, and
$\underline{P}_{\,\underline{m}}=i\underline{\partial}_{\,\underline{m}}$
is the generator of translations on quantum fields
$\underline{\phi}(\eta)\equiv\underline{\phi}(t, q^{1})$. According
to the embedding map~(\ref{S7}), the relation between the fields
$\phi(x)$ and $\underline{\phi}(\eta)$ can be given by the a proper
orthochronous Lorentz transformation. For a field of spin-$\vec{S}$,
the general transformation law reads
\begin{equation}
\begin{array}{l}
\phi'_{\alpha}(x')=M^{\phantom{a}\beta}_{\alpha}\,\phi_{\beta}(x)=
\exp\left(-\frac{1}{2}\theta^{mn}S_{mn}\right)^{\phantom{a}\beta}_{\alpha}\,\phi_{\beta}(x)=
\exp\left(-i\vec{\theta}\cdot
\vec{S}-i\vec{\zeta}\cdot\vec{K}\right)^{\phantom{a}\beta}_{\alpha}\,\phi_{\beta}(x),
\end{array}
\label{S11}
\end{equation}
where $\vec{\theta}$ is the rotation angle about an axis $\vec{n}$
$(\vec{\theta}\equiv \theta\vec{n})$, and $\vec{\zeta}$ is the boost
vector $\vec{\zeta}\equiv \vec{e}_{v}\cdot\tan h^{-1}\,\beta$,
provided $\vec{e}_{v}\equiv \vec{v}/|\vec{v}|$, $\beta \equiv
|\vec{v}|/c$, $\theta^{i}\equiv (1/2)\varepsilon^{ijk}\,\theta_{k}$
($i,j,k=1,2,3$), and $\zeta^{i}\equiv \theta^{i0}=-\theta^{0i}$. The
antisymmetric tensor $S_{mn}=-S_{nm}$, satisfying the commutation
relations of the $SL(2.C)$, is the (finite-dimensional) irreducible
matrix representations of the Lie algebra of the Lorentz group, and
$\alpha$ and $\beta$ label the components of the matrix
representation space, the dimension of which is related to the spin
$S^{i}\equiv (1/2)\varepsilon^{ijk}\,S_{k}$ of the particle. The
spin $\vec{S}$ generates three-dimensional rotations in space and
the $K^{i}\equiv S^{0i}$ generate the Lorentz-boosts. The fields of
spin-zero $(\vec{S}=\vec{K}=0)$ scalar field $A(x)$ and spin-one
$A^{n}(x)$, corresponding to the $(1/2. 1/2)$ representation,
transform under a general Lorentz transformation as
\begin{equation}
\begin{array}{l}
\underline{A}(\eta)=A(x), \quad\mbox{spin}\quad 0;\quad
\underline{A}^{m}(\eta)=\Lambda^{m}_{\phantom{a}n}\,A^{n}(x),
\quad\mbox{spin} \quad 1,
\end{array}
\label{S12}
\end{equation}
where the Lorentz transformation is written as
\begin{equation}
\begin{array}{l}
\Lambda^{m}_{\phantom{a}n}(M)\equiv
\frac{1}{2}\,Tr\left(\sigma_{m}M\sigma_{n}M^{\dag}\right),
\end{array}
\label{S13}
\end{equation}
provided, $\sigma^{m}\equiv (I_{2}, \vec{\sigma})$, $\vec{\sigma}$
are Pauli spin matrices. A two-component $(1/2, 0)$ Weyl fermion
$\chi_{\beta}(x)$ transforms under Lorentz transformation, in accord
to embedding map~(\ref{S7}), as
\begin{equation}
\begin{array}{l}
\chi_{\beta}(x)\,\longrightarrow\,\underline{\chi}_{\,\alpha}(\eta)=(M_{R})^{\phantom{a}\beta}_{\alpha}\,\chi_{\beta}(x),\quad
\alpha,\beta=1,2
\end{array}
\label{S14}
\end{equation}
where the rotation matrix is given as
\begin{equation}
\begin{array}{l}
M_{R}=e^{i\frac{1}{2}\sigma_{2}\theta_{2}}e^{i\frac{1}{2}\sigma_{3}\theta_{3}}.
\end{array}
\label{S15}
\end{equation}
The matrix $M_{R}$ corresponds to the rotation of an hermitian
$2\times 2$ matrix $p^{n}\sigma_{n}$:
\begin{equation}
\begin{array}{l}
p^{m}_{q}\sigma_{m}= M_{R}\,p^{n}\sigma_{n}\, M_{R}^{\dag},
\end{array}
\label{S16}
\end{equation}
by the angles $\theta_{3}$ and $\theta_{2}$ about the axes $n_{3}$
and $n_{2}$, respectively, where the standard momentum is
$p^{n}\equiv m (ch \beta,\, sh \beta \sin \theta_{2} \cos
\theta_{3},\,sh \beta \sin \theta_{2} \sin \theta_{3},\, sh \beta
\cos \theta_{2})$, and $p^{m}_{q}$ is $p^{m}_{q}\equiv m (ch
\beta,\,0,\,0,\,sh \beta)$. According to ~(\ref{S14}), a
two-component $(0, 1/2)$ Weyl spinor field is denoted by
$\bar{\chi}^{\dot{\beta}}(x)$, and transforms as
\begin{equation}
\begin{array}{l}
\bar{\chi}^{\dot{\beta}}(x)\,\longrightarrow\,\underline{\,\bar{\chi}}^{\dot{\alpha}}(\eta)=
(M_{R}^{-1})^{\dag\dot{\alpha}}_{\phantom{ab}\dot{\beta}}\,\bar{\chi}^{\dot{\beta}}(x),\quad
\dot{\alpha},\dot{\beta}=1,2
\end{array}
\label{S17}
\end{equation}
where we have used
$(M^{\dag})^{\dot{\beta}}_{\phantom{a}\dot{\alpha}}=(M^{*})^{\phantom{a}\dot{\beta}}_{\dot{\alpha}}$.
The so-called 'dotted' indices have been introduced to distinguish
the $(0, 1/2)$ representation from the $(1/2, 0)$ representation.
The "bar" over the spinor is a convention that this is the $(0,
1/2)$-representation. The infinitesimal Lorentz transformation
matrices for the $(1/2, 0)$ and $(0, 1/2)$ representations,
\begin{equation}
\begin{array}{l}
 M\simeq I_{2}-\frac{i}{2}\vec{\theta}\cdot
\vec{\sigma}-\frac{1}{2}\vec{\zeta}\cdot\vec{\sigma},
\quad\mbox{for}\, (\frac{1}{2},0);\quad (M^{-1})^{\dag}\simeq
I_{2}-\frac{i}{2}\vec{\theta}\cdot
\vec{\sigma}+\frac{1}{2}\vec{\zeta}\cdot\vec{\sigma},
\quad\mbox{for}\quad (0,\frac{1}{2})
\end{array}
\label{S18}
\end{equation}
give $S^{mn}=\sigma^{mn}$ for the  $(1/2, 0)$ representation and
$S^{mn}=\bar{\sigma}{}^{mn}$ for the $(0, 1/2)$ representation,
where the bilinear covariants that transform as a Lorentz
second-rank tensor read
\begin{equation}
\begin{array}{l}
 (\sigma^{mn})^{\phantom{a} \beta}_{\alpha}\equiv
\frac{i}{4}(\sigma^{m}_{\alpha\dot{\alpha}}\,\bar{\sigma}{}^{n\dot{\alpha}\beta}-
\sigma^{n}_{\alpha\dot{\alpha}}\,\bar{\sigma}{}^{m\dot{\alpha}\beta}),\quad
(\bar{\sigma}{}^{mn})_{\phantom{a} \dot{\beta}}^{\dot{\alpha}}\equiv
\frac{i}{4}(\bar{\sigma}{}^{m\dot{\alpha}\alpha}\,\sigma^{n}_{\alpha\dot{\beta}}-
\bar{\sigma}{}^{n\dot{\alpha}\alpha}\,\sigma^{m}_{\alpha\dot{\beta}}),
\end{array}
\label{S19}
\end{equation}
provided $\bar{\sigma}{}^{m}\equiv (I_{2};\,-\vec{\sigma})$,
$\,(\sigma^{m\,*})_{\alpha\dot{\beta}}=\sigma^{m}_{\beta\dot{\alpha}}$
and
$(\bar{\sigma}{}^{m\,*})^{\dot{\alpha}\beta}=\bar{\sigma}{}^{m\dot{\beta}\alpha}$.

\section{MS-SUSY}
As alluded to above, a {\it creation} of a particle in
$\underline{M}_{\,2}$ means its transition from $M_{4}$ to
$\underline{M}_{\,2}$, while an {\it annihilation} of a particle in
$\underline{M}_{\,2}$ means vice versa. The same interpretation
holds for the {\it creation} and {\it annihilation} processes in
$M_{4}$. Since all fermionic and bosonic states, taken together,
form a basis in the Hilbert space, the basis vectors in the Hilbert
space, therefore, can be written in the form
$|\underline{n}_{\,b},\,n_{f}>$ or $|n_{b},\,\underline{n}_{\,f}>$,
where the boson and  fermion occupation numbers are $n_{b}$ or
$\underline{n}_{\,b}\,(=0,1,...,\infty)$ and $n_{f}$ or
$\underline{n}_{\,f}\,(=0,1)$. So, we may construct the quantum
operators, $(q^{\dag},\, \underline{q}^{\dag})$ and $(q,\,
\underline{q})$, which replace bosons by fermions and fermions by
bosons, respectively,
\begin{equation}
\begin{array}{l}

q^{\dag}\,|\underline{n}_{\,b},\,n_{f}>\,\longrightarrow\,|\underline{n}_{\,b}-1,\,n_{f}+1>,
\quad
q\,|\underline{n}_{\,b},\,n_{f}>\,\longrightarrow\,|\underline{n}_{\,b}+1,\,n_{f}-1>,
\end{array}
\label{S20}
\end{equation}
and that
\begin{equation}
\begin{array}{l}

\underline{q}^{\dag}\,|n_{b},\,\underline{n}_{\,f}>\,\longrightarrow\,|n_{b}-1,\,\underline{n}_{\,f}+1>,
\quad
\underline{q}\,|n_{b},\,\underline{n}_{\,f}>\,\longrightarrow\,|n_{b}+1,\,\underline{n}_{\,f}-1>.
\end{array}
\label{S21}
\end{equation}
This framework combines bosonic and fermionic states on the same
footing, rotating them into each other under the action of operators
$q$ and $\underline{q}$. Consider two pairs of creation and
annihilation operators $(b^{\dag},\,b)$  and $(f^{\dag},\,f)$ for
bosons and fermions, respectively, referred to the background space
$M_{4}$, as well as $(\underline{b}^{\dag},\,\underline{b})$ and
$(\underline{f}^{\dag},\,\underline{f})$ for bosons and fermions,
respectively, related to the background master space
$\underline{M}_{\,2}$. Putting two operators in one $B=
(\underline{b}$  or $ b)$ and $F= (f$ or $\underline{f})$, the
canonical quantization rules can be written most elegantly as
\begin{equation}
\begin{array}{l}
[B,\,B^{\dag}]=1; \quad \{F,\,F^{\dag}\}=1; \quad
[B,\,B]=[B^{\dag},\,B^{\dag}]=\{F,\,F\}=\\\{F^{\dag},\,F^{\dag}\}=
[B,\,F]=[B,\,F^{\dag}]=[B^{\dag},\,F]=[B^{\dag},\,F^{\dag}]=0,
\end{array}
\label{S26}
\end{equation}
where we note that $\delta_{ij}\delta^{3}(\vec{p}-\vec{p}')$ and
$\delta_{ij}\delta^{3}(\vec{p}_{q}-\vec{p}_{q}')$ are the unit
element 1 of the convolution product *, and according to embedding
map~(\ref{S7}) we have $p_{q}=\pm |\vec{p}|$ and $p_{q}'=\pm
|\vec{p}'|$. The operators $q$ and $\underline{q}$ can be
constructed as
\begin{equation}
\begin{array}{l}
q^{\dag}= q_{0}\,\underline{b}\,f^{\dag},\quad
q=q_{0},\underline{b}^{\dag} f,\quad \underline{q}^{\dag}=q_{0}\,
b\,\underline{f}^{\dag},\quad \underline{q}=q_{0}\, b^{\dag}
\underline{f}.
\end{array}
\label{S27}
\end{equation}
So, we may refer the action of the supercharge operators $q$ and
$q^{\dag}$ to the background space $M_{4}$, having applied in the
chain of following transformations of fermion $\chi$ (accompanied
with the auxiliary field $F$ as it will be seen later on) to boson
$\underline{A}$, defined on $\underline{M}_{\,2}$:
\begin{equation}
\begin{array}{l}
\cdots \,\longrightarrow\,
\chi^{(F)}\,\longrightarrow\,\underline{A}\,\longrightarrow\,
\chi^{(F)}\,\longrightarrow\,\underline{A}\,\longrightarrow\,\chi^{(F)}\,\longrightarrow\,\cdots
\end{array}
\label{S33}
\end{equation}
Respectively, we may refer the action of the supercharge operators
$\underline{q}$ and $\underline{q}^{\dag}$ to the
$\underline{M}_{\,2}$, having applied in the chain of following
transformations of fermion $\underline{\chi}$ (accompanied with the
auxiliary field $\underline{F}$) to boson $A$, defined on the
background space $M_{4}$:
\begin{equation}
\begin{array}{l}
\cdots \,\longrightarrow\,
\underline{\chi}^{(\underline{F})}\,\longrightarrow\,A\,\longrightarrow\,
\underline{\chi}^{(\underline{F})}\,\longrightarrow\,A\,\longrightarrow\,
\underline{\chi}^{(\underline{F})}\,\longrightarrow\,\cdots
\end{array}
\label{S34}
\end{equation}
Written in one notation, $Q=(q$ or $\underline{q})$, the operators
~(\ref{S27}) become
\begin{equation}
\begin{array}{l}
Q=q_{0}\,B^{\dag} F=(q\,\, \mbox{or}\,\, \underline{q}),\quad
Q^{\dag}=q_{0}\,B F^{\dag}=(q^{\dag}\,\, \mbox{or}\,\,
\underline{q}^{\dag}).
\end{array}
\label{S28}
\end{equation}
Due to nilpotent fermionic operators $F^{2}=(F^{\dag})^{2}=0$, the
operators $Q$ and $Q^{}\dag$ also are nilpotent: $
Q^{2}=(Q^{\dag})^{2}=0 . $ Hence, the quantum system can be
described in one notation by the selfadjoint Hamiltonian ${\cal
H}=(H_{q}\equiv\{q^{\dag},\,q\}$ or
$H_{\underline{q}}\equiv\{\underline{q}^{\dag},\,\underline{q}\})$,
and that the generators $Q$ and $Q^\dag$ satisfy an algebra of
anticommutation and commutation relations:
\begin{equation}
\begin{array}{l}
{\cal H}=\{Q^{\dag},\,Q\}\geq 0;\quad [{\cal H},\,Q]=[{\cal
H},\,Q^{\dag}]=0.
\end{array}
\label{S29}
\end{equation}
This is a sum of Hamiltonian of bosonic and fermionic noninteracting
oscillators, which decouples, for $Q=q$, into
\begin{equation}
\begin{array}{l}

H_{q}=q_{0}^{2}\,(\underline{b}^{\dag}\underline{b}+f^{\dag}f)=q_{0}^{2}\,(\underline{b}^{\dag}\underline{b}+\frac{1}{2})
+q_{0}^{2}\,(f^{\dag}f-\frac{1}{2})\equiv H_{\underline{b}}+H_{f},
\end{array}
\label{SS301}
\end{equation}
or, for $Q=\underline{q}$, into
\begin{equation}
\begin{array}{l}

H_{\underline{q}}=q_{0}^{2}\,(b^{\dag}b+\underline{f}^{\dag}\underline{f})=
q_{0}^{2}\,(b^{\dag}b+\frac{1}{2})
+q_{0}^{2}\,(\underline{f}^{\dag}\underline{f}-\frac{1}{2})\equiv
H_{b}+H_{\underline{f}},
\end{array}
\label{SS302}
\end{equation}
with the corresponding energies:
\begin{equation}
\begin{array}{l}

E_{q}=q_{0}^{2}\,(\underline{n}_{b}+\frac{1}{2})+q_{0}^{2}\,(n_{f}-\frac{1}{2}),\quad
E_{\underline{q}}=q_{0}^{2}\,(n_{b}+\frac{1}{2})+q_{0}^{2}\,(\underline{n}_{f}-\frac{1}{2}).
\end{array}
\label{S31}
\end{equation}
This formalism manifests its practical and technical virtue in the
proposed algebra~(\ref{S29}), which  becomes more clear in a
normalization $q_{0}=\sqrt{m}$:
\begin{equation}
\begin{array}{l}
\{Q^{\dag},\,Q\}=2m;\quad \{Q,\,Q\}=\{Q^{\dag},\,Q^{\dag}\}=0.
\end{array}
\label{S32}
\end{equation}
The latter has underlying algebraic structure of the superalgebra
for massive {\it one-particle} states in the rest frame of  $N=1$
SUSY theory without central charges, see e.g.~\cite{DHM}-\cite{DM}.
This is rather technical topic, and it requires care to do
correctly. In what follows we only give a brief sketch. The
extension of the MS-SUSY superalgebra~(\ref{S32}) in general case
when $\vec{p}=i\vec{\partial}\neq 0$ in $M_{4}$ or
$p_{q}=i\partial_{q}\neq 0$ in $\underline{M}_{\,2}$, and assuming
that the resulting motion of a particle in $M_{4}$ is governed by
the Lorentz symmetries, the MS-SUSY algebra can then be summarized
as a non-trivial extension of the Poincar\'{e} group algebra thus of
the commutation relations of the bosonic generators of four momenta
and six Lorentz generators referred to $M_{4}$. Moreover, if there
are several spinor generators $Q^{\phantom{a}i}_{\alpha}$ with
$i=1,...,N$ - theory with $N-$extended supersymmetry, can be written
as a graded Lie algebra (GLA) of SUSY field theories, with commuting
and anticommuting generators:
\begin{equation}
\begin{array}{l}
\{Q^{\phantom{a}i}_{\alpha},\,\bar{Q}{}^{j}_{\phantom{a}\dot{\alpha}}\}=2\delta^{ij}\,\sigma^{\hat{m}}_{\alpha\dot{\alpha}}\,
p_{\hat{m}};\quad
[p_{\hat{m}},\,Q^{\phantom{a}i}_{\alpha}]=[p_{\hat{m}},\,\bar{Q}{}^{j}_{\phantom{a}\dot{\alpha}}]=0,\\
\{Q^{\phantom{a}i}_{\alpha},\,Q^{\phantom{a}j}_{\beta}\}=\{\bar{Q}{}^{i}_{\phantom{a}\dot{\alpha}},\,
\bar{Q}{}^{j}_{\phantom{a}\dot{\beta}}\}=0; \quad [p_{\hat{m}},\,
p_{\hat{n}}]=0.
\end{array}
\label{S36}
\end{equation}
Here $\sigma^{(\pm)}=(1/2)(\sigma^{o}\pm \sigma^{3})$, and in order
to trace a maximal resemblance in outward appearance to the standard
SUSY theories, we set one notation $\hat{m}= (m \quad\mbox{if}\quad
Q=q,$ \,or\, $\underline{m}\quad\mbox{if}\quad Q=\underline{q})$, no
sum over $\hat{m}$,  and as before the indices $\alpha$ and
$\dot{\alpha}$ go over $1$ and $2$. So for both supercharges, $q$
and $\underline{q}$, we get a supersymmetric models, respectively:
\begin{equation}
\begin{array}{l}
\{q^{\phantom{a}i}_{\alpha},\,\bar{q}{}^{j}_{\phantom{a}\dot{\alpha}}\}=2\delta^{ij}\,\sigma^{m}_{\alpha\dot{\alpha}}\,
p_{m};\quad
[p_{m},\,q^{\phantom{a}i}_{\alpha}]=[p_{m},\,\bar{q}{}^{j}_{\phantom{a}\dot{\alpha}}]=0,\\
\{q^{\phantom{a}i}_{\alpha},\,q^{\phantom{a}j}_{\beta}\}=\{\bar{q}{}^{i}_{\phantom{a}\dot{\alpha}},\,
\bar{q}{}^{j}_{\phantom{a}\dot{\beta}}\}=0; \quad [p_{m},\,
p_{n}]=0.
\end{array}
\label{SS361}
\end{equation}
and
\begin{equation}
\begin{array}{l}
\{\underline{q}^{\phantom{a}i}_{\,\alpha},\,\bar{\underline{q}}{}^{j}_{\phantom{a}\dot{\alpha}}\}=
2\delta^{ij}\,\sigma^{\underline{m}}_{\alpha\dot{\alpha}}\,
p_{\underline{m}};\quad
[p_{\underline{m}},\,\underline{q}^{\phantom{a}i}_{\,\alpha}]=[p_{\underline{m}},\,
\bar{\underline{q}}{}^{j}_{\phantom{a}\dot{\alpha}}]=0,\\
\{\underline{q}^{\phantom{a}i}_{\,\alpha},\,\underline{q}^{\phantom{a}j}_{\,\beta}\}=
\{\bar{\underline{q}}{}^{i}_{\phantom{a}\dot{\alpha}},\,
\bar{\underline{q}}{}^{j}_{\phantom{a}\dot{\beta}}\}=0; \quad
[p_{\underline{m}},\, p_{\underline{n}}]=0.
\end{array}
\label{SS362}
\end{equation}
For the self-contained arguments, we should emphasize  the crucial
differences between the MS-induced SUSY and
the standard theories as follows: \\
1) The standard theory can be realized only as a spontaneously
broken symmetry since the experiments do not show elementary
particles to be accompanied by superpartners with different spin but
identical mass. The MS-SUSY, in contrary, can only be realized as an
{\it unbroken supersymmetry}. \\
2) In the standard theory, the Q's operate on the fields defined on
the single $M_{4}$ space. It is why the result of a Lorentz
transformation in $M_{4}$ followed by a supersymmetry transformation
is different from that when the order of the transformations is
reversed~\cite{Soh}. But, in the MS-SUSY theory, the Q's
((\ref{S27}), (\ref{S28})) operate on the fields defined on both
$M_{4}$ and $\underline{M}_{\,2}$ spaces, fulfilling a transition of
a particle between these spaces ($M_{4}\rightleftharpoons
\underline{M}_{\,2}$).  So after a Lorentz transformation in $M_{4}$
followed by a supersymmetry transformation (which, as we shall see
below, now results to uniform motion of a particle with initial
constant velocity) we have a particle moving with changed constant
velocity. We obtain the same result if we reverse the order of the
transformations, namely a Lorentz transformation changes the initial
velocity and a supersymmetry transformation followed by a Lorentz
transformation just keep the uniform motion with the changed
velocity.

We shall forbear to write out further the unitary representations of
supersymmetry, giving rise to the notion of supermultiplets, as they
are so well known. Also, unless otherwise stated we will not discuss
the theories with $N > 1$, because  it is unlikely that they play
any role in low-energy physics.

\subsection{Wess-Zumino model}
To obtain a feeling for this model we may consider first example of
non-trivial linear representation of the MS-SUSY algebra in analogy
of the Wess-Zumino toy model \cite{WZ}, which has $N=1$ and
$s_{0}=0,$ and contains two spin states of a massive Majorana spinor
$\psi (\chi,\,\underline{\chi})$ and two complex scalar fields
${\cal A}(A,\, \underline{A})$ and auxiliary fields ${\cal F}(F,\,
\underline{F})$, which provide in supersymmetry theory the fermionic
and bosonic degrees of freedom to be equal. This model is
instructive because it contains the essential elements of the
MS-induced SUSY. Let us first introduce four additional,
anticommuting (Grassmann) parameters $\epsilon^{\alpha}
(\xi^{\alpha},\,\underline{\xi}^{\alpha})$ and
$\bar{\epsilon}{}^{\alpha}
(\bar{\xi}{}^{\alpha},\,\bar{\underline{\xi}}^{\alpha})$:
\begin{equation}
\begin{array}{l}
\{\epsilon^{\alpha},\,\epsilon^{\beta}\}=\{\bar{\epsilon}{}^{\alpha},\,\bar{\epsilon}{}^{\beta}\}=
\{\epsilon^{\alpha},\,\bar{\epsilon}{}^{\beta}\}=0,\quad
\{\epsilon^{\alpha},\,Q_{\beta}\}=\dots=[p_{\hat{m}},\,\epsilon^{\alpha}]=0,
\end{array}
\label{S46}
\end{equation}
which allow to write the algebra~(\ref{S36}) $(N=1)$ in terms of
commutators only:
\begin{equation}
\begin{array}{l}
 [\epsilon Q,\,\bar{Q}\bar{\epsilon}
]=2\epsilon\sigma^{\hat{m}}\bar{\epsilon}p_{\hat{m}},\quad [\epsilon
Q,\,\epsilon Q]=[ \bar{Q}\bar{\epsilon},\,
\bar{Q}\bar{\epsilon}]=[p^{\hat{m}},\,\epsilon Q]=[p^{\hat{m}},\,
\bar{Q}\bar{\epsilon}]=0.
\end{array}
\label{S47}
\end{equation}
Here we have dropped the indices $\epsilon Q=\epsilon^{\alpha}
Q_{\alpha}$ and $\bar{\epsilon}
\bar{Q}=\bar{\epsilon}_{\dot{\alpha}} \bar{Q}{}^{\dot{\alpha}}$. The
infinitesimal supersymmetry transformations for $Q=q$ read
\begin{equation}
\begin{array}{l}
\delta_{\xi}\underline{A}=(\xi q + \bar{\xi}
\bar{q})\times \underline{A}=\sqrt{2} \xi\chi,\\
\delta_{\xi}\chi=(\xi q + \bar{\xi}
\bar{q})\times\chi=i\sqrt{2}\sigma^{m}\bar{\xi}\partial_{m}\underline{A}+\sqrt{2}\xi F,\\
\delta_{\xi}F=(\xi q + \bar{\xi} \bar{q})\times
F=i\sqrt{2}\bar{\xi}\bar{\sigma}{}^{m}\partial_{m}\chi;
\end{array}
\label{S49}
\end{equation}
and for $Q=\underline{q}$ are in the form
\begin{equation}
\begin{array}{l}
\delta_{\,\underline{\xi}}A=(\underline{\xi} \,\underline{q} +
\bar{\underline{\xi}}\,
\bar{\underline{q}})\times A=\sqrt{2} \,\underline{\xi}\,\underline{\chi},\\
\delta_{\,\underline{\xi}}\,\underline{\chi}=(\underline{\xi}\,
\underline{q} + \bar{\underline{\xi}}\,
\bar{\underline{q}})\times\underline{\chi}=
i\sqrt{2}\,\sigma^{\underline{m}}\,\bar{\underline{\xi}}\,\partial_{\underline{m}}A+\sqrt{2}\underline{\xi}\, \underline{F},\\
\delta_{\,\underline{\xi}}\,\underline{F}=(\underline{\xi}\,
\underline{q} + \bar{\underline{\xi}}\,\bar{\underline{q}})\times
\underline{F}=i\sqrt{2}\,\bar{\underline{\xi}}\,\bar{\sigma}{}^{\underline{m}}\,\partial_{\underline{m}}\,\underline{\chi},
\end{array}
\label{S50}
\end{equation}
where according to ~(\ref{S12}) $A=\underline{A}$. The first
relation in~(\ref{S36}) means that there should be a particular way
of going from one subspace (bosonic/fermionic) to the other and
back, such that the net result is as if we had operator of
translation $p_{\hat{m}}$ on the original subspace. Actually, it can
be checked that the supersymmetry transformations close
supersymmetry algebra:
\begin{equation}
\begin{array}{l}
[\delta_{\epsilon_{1}},\, \delta_{\epsilon_{2}}]{\cal
A}=-2i(\epsilon_{1}\sigma^{\hat{m}}\bar{\epsilon}_{2}-
\epsilon_{2}\sigma^{\hat{m}}\bar{\epsilon}_{1})\,\partial_{\hat{m}}{\cal
A},
\end{array}
\label{S51}
\end{equation}
and likewise for $\psi$ and ${\cal F}$. In the framework of MS-SUSY
theory, the Wess-Zumino model has the following Lagrangians:
\begin{equation}
\begin{array}{l}
{\cal L}_{Q=q}={\cal L}_{0}+m{\cal L}_{m},\quad {\cal
L}_{Q=\underline{q}}=\underline{{\cal L}}_{0}+m\underline{{\cal
L}}_{m},
\end{array}
\label{S52}
\end{equation}
provided,
\begin{equation}
\begin{array}{l}
 {\cal L}_{0}=i\partial_{m}\bar{\chi}\bar{\sigma}^{m}\chi+
\underline{A}^{*}\underline{\square}\,\underline{A}+F^{*}F,\quad
{\cal L}_{m}=
\underline{A}\,F+\underline{A}^{*}\,F^{*}-\frac{1}{2}\chi\chi-\frac{1}{2}\bar{\chi}\bar{\chi},\\
 \underline{{\cal
L}}_{0}=i\partial_{\underline{m}}\,\bar{\underline{\chi}}\bar{\sigma}^{m}\underline{\chi}+
A^{*}\square\,A+\underline{F}^{*}\underline{F},\quad
\underline{{\cal L}}_{\,\underline{m}}=
A\,\underline{F}+A^{*}\,\underline{F}^{*}-\frac{1}{2}\underline{\chi}\,\underline{\chi}
-\frac{1}{2}\bar{\underline{\chi}}\,\bar{\underline{\chi}},
\end{array}
\label{S54}
\end{equation}
where according to the embedding map~(\ref{S7}), $\square
=\underline{\square}$ and $A=\underline{A}$. Whereupon, the
equations of motion for the Weyl spinor $\psi$ and complex scalar
${\cal A}$ of the same mass $m$, are
\begin{equation}
\begin{array}{l}
i\bar{\sigma}^{m}\,\partial_{m}\,\chi+m\bar{\chi}=0,\qquad\qquad\qquad
i\bar{\sigma}^{\underline{m}}\,\partial_{\underline{m}}\,\underline{\chi}+m\bar{\underline{\chi}}=0,\\
F+m\underline{A}^{*}=0,\qquad\qquad\mbox{or}\qquad
\qquad\underline{F}+m
A^{*}=0,\\
\square\,\underline{A}+m F^{*}=0,\qquad\qquad\qquad\qquad
\underline{\square}\,A+m \underline{F}^{*}=0.\\
\qquad\quad (a) \qquad\qquad\qquad\qquad\qquad\qquad(b)\\
\end{array}
\label{S55}
\end{equation}
By virtue of~(\ref{S33}) and ~(\ref{S34}), respectively, (a) stands
for $Q=q$ (referring to the motion of a fermion, $\chi$, in $M_{4}$)
and (b) stands for $Q=\underline{q}$ (so, of a boson, $A$, in
$M_{4}$). Finally, the algebraic auxiliary field ${\cal F}$ can be
eliminated to find
\begin{equation}
\begin{array}{l}
{\cal
L}_{Q=q}=i\partial_{m}\bar{\chi}\bar{\sigma}^{m}\chi-\frac{1}{2}(\chi\chi+\bar{\chi}\bar{\chi})+
\underline{A}^{*}\underline{\square}\,\underline{A}-m^{2}\underline{A}^{*}\underline{A},\\
\underline{{\cal
L}}_{Q=\underline{q}}=i\partial_{\underline{m}}\,\bar{\underline{\chi}}\bar{\sigma}^{m}\underline{\chi}-
\frac{1}{2}(\underline{\chi}\,\underline{\chi}
+\bar{\underline{\chi}}\,\bar{\underline{\chi}}) + A^{*}\square\,A
-m^{2}A^{*}\,A.
\end{array}
\label{S56}
\end{equation}

\subsection{General superfields}
In the framework of standard generalization of the coset
construction~\cite{W1}-\cite{SS}, we will take $G=G_{q}\times
G_{\underline{q}}$ to be the supergroup generated by the MS-SUSY
algebra~(\ref{S36}). Let the stability group $H=H_{q}\times
H_{\underline{q}}$ be the Lorentz group (as to $M_{4}$ and
$\underline{M}_{\,2}$), and we choose to keep all of $G$ unbroken.
Given $G$ and $H$, we can construct the coset, $G/H$,  by an
equivalence relation on the elements of $G$: $\quad \Omega\sim
\Omega h$, where $\Omega=\Omega_{q}\times \Omega_{\underline{q}}\in
G$ and $h=h_{q}\times h_{\underline{q}}\in H$, so that the coset can
be pictured as a section of a fiber bundle with total space, $G$,
and fiber, $H$. So, the Maurer-Cartan form, $\Omega^{-1}d\Omega$, is
valued in the Lie algebra of $G$, and transforms as follows under a
rigid G transformation,
\begin{equation}
\begin{array}{l}
\Omega\,\longrightarrow\,g\Omega h^{-1},\quad
\Omega^{-1}d\Omega\,\longrightarrow\,h(\Omega^{-1}d\Omega)h^{-1}-dh\,h^{-1},
\end{array}
\label{S57}
\end{equation}
with $g\in G$. Also we consider a superspace which is an enlargement
of $M_{4}\oplus\underline{M}_{\,2}$ (spanned by the coordinates
$X^{\hat{m}}=(x^{m},\,\eta^{\,\underline{m}})$ by the inclusion of
additional fermion coordinates  $\Theta^{\alpha}=
(\theta^{\alpha},\,\underline{\theta}^{\alpha})$ and
$\bar{\Theta}_{\dot{\alpha}}=
(\bar{\theta}_{\dot{\alpha}},\,\underline{\bar{\theta}}_{\dot{\,\alpha}})$,
as to ($q,\,\underline{q}$), respectively.  But note that the
relation between the two spinors $\theta$ and $\underline{\theta}$
should be derived further from the embedding map~(\ref{S7}) (see
next section). These spinors satisfy the following relations:
\begin{equation}
\begin{array}{l}
\{\Theta^{\alpha},\,\Theta^{\beta}\}=\{\bar{\Theta}_{\dot{\alpha}},\,\bar{\Theta}_{\dot{\beta}}\}=
\{\Theta^{\alpha},\,\bar{\Theta}_{\dot{\beta}}\}=0,\\
{}[x^{m},\,\theta^{\alpha}]=
[x^{m},\,\bar{\theta}_{\dot{\alpha}}]=0, \quad
[\eta^{\,\underline{m}},\,\underline{\theta}^{\alpha}]=[\eta^{\,\underline{m}},\,\underline{\bar{\theta}}_{\dot{\alpha}}]=0.
\end{array}
\label{S59}
\end{equation}
and ${\Theta^{\alpha}}^{*}=\bar{\Theta}{}^{\dot{\alpha}}$. Points in
superspace are then identified by the generalized coordinates
$z^{M}=(X^{\hat{m}},\,\Theta^{\alpha},\,\bar{\Theta}_{\dot{\alpha}})$.
In case at hand we have then
\begin{equation}
\begin{array}{l}
\Omega
(X,\,\Theta,\,\bar{\Theta})=e^{i(-X^{\hat{m}}p_{\hat{m}}+\Theta^{\alpha}Q_{\alpha}+
\bar{\Theta}_{\dot{\alpha}}\bar{Q}{}^{\dot{\alpha}})}=
\Omega_{q}(x,\,\theta,\,\bar{{\theta}})\times
\Omega_{\underline{q}}(\eta,\,\underline{\theta},\,\bar{\underline{\theta}}),
\end{array}
\label{S58}
\end{equation}
where we now imply a summation over $\hat{m}$, etc., such that
\begin{equation}
\begin{array}{l}
\Omega_{q}(x,\,\theta,\,\bar{{\theta}})=e^{i(-x^{m}p_{m}+\theta^{\alpha}q_{\alpha}+
\bar{\theta}_{\dot{\alpha}}\bar{q}{}^{\dot{\alpha}})},\quad
\Omega_{\underline{q}}(\eta,\,\underline{\theta},\,\bar{\underline{\theta}})=
e^{i(-\eta^{\underline{m}}p_{\underline{m}}+\underline{\theta}^{\alpha}\underline{q}_{\,\alpha}+
\bar{\underline{\theta}}_{\dot{\,\alpha}}\bar{\underline{q}}{}^{\dot{\alpha}})}.
\end{array}
\label{SS58}
\end{equation}
Supersymmetry transformation will be defined as a translation in
superspace, specified by the group element
\begin{equation}
\begin{array}{l}
g(0,\,\epsilon,\,\bar{\epsilon})=e^{i(\epsilon^{\alpha}Q_{\alpha}+\bar{\epsilon}_{\dot{\alpha}}\bar{Q}^{\dot{\alpha}})}=
g_{q}(0,\,\xi,\,\bar{\xi})\times
g_{\underline{q}}(0,\,\underline{\xi},\,\bar{\underline{\xi}})=
e^{i(\xi^{\alpha}q_{\alpha}+\bar{\xi}_{\dot{\alpha}}\bar{q}^{\dot{\alpha}})}\times
e^{i(\underline{\xi}^{\alpha}\underline{q}_{\,\alpha}+\bar{\underline{\xi}}_{\,\dot{\alpha}}\bar{\underline{q}}^{\dot{\alpha}})},
\end{array}
\label{SSS58}
\end{equation}
with corresponding anticommuting parameters  $\epsilon=(\xi$ or
$\underline{\xi}$). To study the effect of supersymmetry
transformations~(\ref{S57}) and $h=1$, we consider
\begin{equation}
\begin{array}{l}
g(0,\,\epsilon,\,\bar{\epsilon})\,\Omega
(X,\,\Theta,\,\bar{\Theta})=e^{i(\epsilon^{\alpha}Q_{\alpha}+\bar{\epsilon}_{\dot{\alpha}}\bar{Q}^{\dot{\alpha}})}\,
e^{i(-X^{\hat{m}}p_{\hat{m}}+\Theta^{\alpha}Q_{\alpha}+
\bar{\Theta}_{\dot{\alpha}}\bar{Q}{}^{\dot{\alpha}})}.
\end{array}
\label{S60}
\end{equation}
The multiplication of two successive transformations can be computed
with the help of the Baker-Campbell-Hausdorf formula
$e^{A}e^{B}=e^{A+B+(1/2)[A,\,B]+\cdots}$. Hence the transformation
~(\ref{S60}) induces the motion
\begin{equation}
\begin{array}{l}
 g(0,\,\epsilon,\,\bar{\epsilon})\,\Omega
(X^{\hat{m}},\,\Theta,\,\bar{\Theta})\,\longrightarrow\,
(X^{\hat{m}}+i\,\Theta\,\sigma^{\hat{m}}\,\bar{\epsilon}-i\,\epsilon\,\sigma^{\hat{m}}\,\bar{\Theta},\,\,
\Theta+\epsilon,\,\bar{\Theta}+\bar{\epsilon}),
\end{array}
\label{S61}
\end{equation}
namely,
\begin{equation}
\begin{array}{l}

g_{q}(0,\,\xi,\,\bar{\xi})\,\Omega_{q}(x,\,\theta,\,\bar{{\theta}})\,\longrightarrow\,
(x^{m}+i\,\theta\,\sigma^{m}\,\bar{\xi}-i\,\xi\,\sigma^{m}\,\bar{\theta},\,\,
\theta+\xi,\,\bar{\theta}+\bar{\xi}),\\

g_{\underline{q}}(0,\,\underline{\xi},\,\bar{\underline{\xi}})\,\Omega_{\underline{q}}(\eta,\,\underline{\theta},\,
\bar{\underline{\theta}})\,\longrightarrow\,
(\eta^{\underline{m}}+i\,\underline{\theta}\,\sigma^{\underline{m}}\,\bar{\underline{\xi}}-
i\,\underline{\xi}\,\sigma^{\underline{m}}\,\bar{\underline{\theta}},\,\,
\underline{\theta}+\underline{\xi},\,\bar{\underline{\theta}}+\bar{\underline{\xi}}).
\end{array}
\label{SS61}
\end{equation}
The superfield $\Phi(z^{M})$, which has a finite number of terms in
its expansion in terms of $\Theta$ and $\bar{\Theta}$ owing to their
anticommuting property, can be considered as the generator of the
various components of the supermultiplets. We will consider only a
scalar superfield $\Phi'({z^{M}}')=\Phi(z^{M})$, an infinitesimal
supersymmetry transformation of which is given as
\begin{equation}
\begin{array}{l}
\delta_{\epsilon}\,\Phi(z^{M})=(\epsilon^{\alpha}Q_{\alpha}+\bar{\epsilon}_{\dot{\alpha}}\bar{Q}^{\dot{\alpha}})\times
\Phi(z^{M}).
\end{array}
\label{S62}
\end{equation}
Acting on this space of functions, the $Q$ and $\bar{Q}$  can be
represented as differential operators:
\begin{equation}
\begin{array}{l}
Q_{\alpha}=\frac{\partial}{\partial
\Theta^{\alpha}}-i\sigma^{\hat{m}}_{\phantom{a}\alpha\dot{\alpha}}\bar{\Theta}{}^{\dot{\alpha}}\partial_{\hat{m}},\quad
\bar{Q}^{\dot{\alpha}}=\frac{\partial}{\partial
\bar{\Theta}_{\dot{\alpha}}}-i\Theta^{\alpha}\sigma^{\hat{m}}_{\phantom{a}\alpha\dot{\beta}}
\varepsilon^{\dot{\beta}\dot{\alpha}}\partial_{\hat{m}},
\end{array}
\label{S63}
\end{equation}
where, as usual, the undotted/dotted spinor indices can be raised
and lowered with a two dimensional undotted/dotted
$\varepsilon-$tensors, and the anticommuting derivatives obey the
relations
\begin{equation}
\begin{array}{l}
\frac{\partial}{\partial
\Theta^{\alpha}}\,\Theta^{\beta}=\delta^{\beta}_{\alpha}, \quad
\frac{\partial}{\partial
\Theta^{\alpha}}\,\Theta^{\beta}\Theta^{\gamma}=\delta^{\beta}_{\alpha}\Theta^{\gamma}-\delta^{\gamma}_{\alpha}\Theta^{\beta},
\end{array}
\label{S64}
\end{equation}
and similarly for $\bar{\Theta}$. In order to write the exterior
product in terms of differential operators, one induces a new basis
as
\begin{equation}
\begin{array}{l}
e^{A}(z)=dZ^{M}\,e^{\phantom{a} A}_{M}(z),
\end{array}
\label{S65}
\end{equation}
and that
\begin{equation}
\begin{array}{l}
D_{A}=e^{\phantom{a} N}_{A}(z)\frac{\partial}{\partial z^{N}},
\end{array}
\label{S66}
\end{equation}
where to be brief we left implicit the symbol $\wedge$ in writing of
exterior product. The covariant derivative operators
\begin{equation}
\begin{array}{l}
 D_{\hat{m}}=\partial_{\hat{m}},\quad
D_{\alpha}=\frac{\partial}{\partial
\Theta^{\alpha}}+i\sigma^{\hat{m}}_{\phantom{a}\alpha\dot{\alpha}}\bar{\Theta}{}^{\dot{\alpha}}\partial_{\hat{m}},\quad
\bar{D}^{\dot{\alpha}}=\frac{\partial}{\partial
\bar{\Theta}_{\dot{\alpha}}}+i\Theta^{\alpha}\sigma^{\hat{m}}_{\phantom{a}\alpha\dot{\beta}}
\varepsilon^{\dot{\beta}\dot{\alpha}}\partial_{\hat{m}},
\end{array}
\label{S67}
\end{equation}
anticommute with the $Q$ and $\bar{Q}$
\begin{equation}
\begin{array}{l}
\{Q_{\alpha},\,D_{\beta}\}=\{\bar{Q}_{\dot{\alpha}},\,\bar{D}_{\dot{\beta}}\}=
\{Q_{\alpha},\,\bar{D}_{\dot{\beta}}\}=\{\bar{Q}_{\dot{\alpha}},\,D_{\beta}\}=0,
\end{array}
\label{S68}
\end{equation}
and satisfy the following structure relations:
\begin{equation}
\begin{array}{l}
\{D_{\alpha},\,D_{\dot{\alpha}}\}=-2i\sigma^{\hat{m}}_{\phantom{a}\alpha\dot{\alpha}}\partial_{\hat{m}},\quad
\{D_{\alpha},\,D_{\beta}\}=\{\bar{D}_{\dot{\alpha}},\,\bar{D}_{\dot{\beta}}\}=0.
\end{array}
\label{S69}
\end{equation}
From~(\ref{S67}), we obtain
\begin{equation}
\begin{array}{l}
e^{\phantom{a} M}_{A}=\left(
                        \begin{array}{cll}
                          e^{\phantom{a} \hat{m}}_{\hat{a}}=\delta^{\hat{m}}_{\hat{a}} &\, e^{\phantom{a} \mu}_{\hat{a}}=0
                          &\, e_{\hat{a}\,\dot{\mu}}=0 \\
                          e^{\phantom{a} \hat{m}}_{\alpha}=i\sigma^{\hat{m}}_{\phantom{a}\alpha\dot{\alpha}}\bar{\Theta}{}^{\dot{\alpha}}
                          &\,
                          e^{\phantom{a}\mu}_{\alpha}=\delta^{\mu}_{\alpha} &\, e_{\alpha\,\dot{\mu}}=0 \\
                          e^{\dot{\alpha}\, \hat{m}}=
                          i\Theta^{\alpha}\sigma^{\hat{m}}_{\phantom{a}\alpha\dot{\beta}}\varepsilon^{\dot{\beta}\dot{\alpha}}
                          &\, e^{\dot{\alpha}\,\mu}=0 &\, e^{\dot{\alpha}}_{\phantom{a}\dot{\mu}}=\delta^{\dot{\alpha}}_{\dot{\mu}} \\
                        \end{array}
                      \right),

\end{array}
\label{SS70}
\end{equation}
where $\hat{a}=(a$ or $\underline{a}),\quad a=0,1,2,3;\quad
\underline{a}=(+),(-)$. The supersymmetry transformations of the
component fields can be found using the differential
operators~(\ref{S67}). The covariant constraint
\begin{equation}
\begin{array}{l}
\bar{D}_{\dot{\alpha}}\Phi(z^{M})=0,
\end{array}
\label{S70}
\end{equation}
which does not impose equations of motion on the component fields,
defines the chiral superfield, $\Phi$. Under the supersymmetry
transformation~(\ref{S61})  the chiral field transforms as follows:
\begin{equation}
\begin{array}{l}
\delta_{\xi}\Phi=(\xi q+\bar{\xi}\bar{q})\times
\Phi=\delta_{\xi}\,\underline{A}(\eta)+\sqrt{2}\theta\delta_{\xi}\chi(x)+\theta\theta\delta_{\xi}
F(x)+\cdots
\end{array}
\label{S71}
\end{equation}
in case of $Q=q$, and
\begin{equation}
\begin{array}{l}
\delta_{\,\underline{\xi}}\,\underline{\Phi}=(\underline{\xi}
\underline{q}+\underline{\bar{\xi}}\,\underline{\bar{q}})\times
\underline{\Phi}=\delta_{\,\underline{\xi}}\,A(x)+\sqrt{2}\underline{\theta}\delta_{\,\underline{\xi}}\,\underline{\chi}(\eta)+
\underline{\theta}\,\underline{\theta}\delta_{\,\underline{\xi}}
\,\underline{F}(\eta)+\cdots
\end{array}
\label{S72}
\end{equation}
in case of $Q=\underline{q}$, where as before
$A(x)=\underline{A}(\underline{\eta})$, and the supersymmetry
transformations are decoupled to~(\ref{S49}) and ~(\ref{S50}).
Equations~(\ref{S71}) and ~(\ref{S72}) show that the chiral
superfield  contains the same component fields as the Wess-Zumino
model for MS-SUSY theory. The supervolume integrals of products of
superfields constructed in the superspace
$(x^{m},\,\theta,\,\bar{\theta})$ will lead to the supersymmetric
kinetic energy for the Wess-Zumino model
\begin{equation}
\begin{array}{l}
\int\,d^{4}x\,d^{4}\theta\,\Phi^{\dag}\Phi,
\end{array}
\label{S73}
\end{equation}
where the superspace Lagrangian reads
\begin{equation}
\begin{array}{l}
\Phi^{\dag}\Phi=\underline{A}^{*}\,\underline{A}+\cdots
+\theta\theta\bar{\theta}\bar{\theta}[\frac{1}{4}\underline{A}^{*}\,\underline{\square}\,\underline{A}+
\frac{1}{4}\underline{\square}\,\underline{A}^{*}\,\underline{A}-
\frac{1}{2}\partial_{\,\underline{m}}\underline{A}^{*}\,\partial^{\,\underline{m}}\underline{A}+\\
F^{*}F+\frac{i}{2}\partial_{m}\bar{\chi}\bar{\sigma}{}^{m}\chi-
\frac{i}{2}\bar{\chi}\bar{\sigma}{}^{m}\partial_{m}\chi],
\end{array}
\label{S74}
\end{equation}
where $\square A=\underline{\square}\,\underline{A}$ and
$\partial_{\,\underline{m}}\underline{A}^{*}\,\partial^{\,\underline{m}}\underline{A}=
\partial_{m}A^{*}\,\partial^{m}A$. Similarly, the supersymmetric
kinetic energy for the Wess-Zumino model constructed in the
superspace
$(\eta^{\,\underline{m}},\,\underline{\theta},\,\underline{\bar{\theta}})$
for MS-SUSY theory is
\begin{equation}
\begin{array}{l}
\int\,d^{2}\eta\,d^{4}\underline{\theta}\,\underline{\Phi}^{\dag}\,\underline{\Phi},
\end{array}
\label{S75}
\end{equation}
where the superspace Lagrangian is written down
\begin{equation}
\begin{array}{l}
\underline{\Phi}^{\dag}\,\underline{\Phi}=A^{*}A+\cdots
+\underline{\theta}\,\underline{\theta}\,\underline{\bar{\theta}}\,\underline{\bar{\theta}}
[\frac{1}{4}A^{*}\,\square\,A+ \frac{1}{4}\square\,A^{*}A-
\frac{1}{2}\partial_{m}A^{*}\,\partial^{m}A+\\\underline{F}^{*}\,\underline{F}+\frac{i}{2}\partial_{\,\underline{m}}\,\underline{\bar{\chi}}
\,\bar{\sigma}{}^{\,\underline{m}}\,\underline{\chi}-
\frac{i}{2}\,\underline{\bar{\chi}}\bar{\sigma}{}^{\,\underline{m}}\,\partial_{\,\underline{m}}\,\underline{\chi}].
\end{array}
\label{S76}
\end{equation}
To complete the model, we also need superspace expressions for the
masses and couplings, which can be easily found in analogy of the
standard theory, namely: 1) fermion masses and Yukawa couplings,
$(\partial^{2}{\cal P}/\partial {\cal A}^{2})\psi\psi$; and 2) the
scalar potential, ${\cal V}({\cal A},\, {\cal A}^{*})=|\partial
{\cal P}/\partial {\cal A}|^{2}$; where ${\cal
P}=(1/2)\,m\,\Phi^{2}+(1/3)\,\lambda\,\Phi^{3}$ is the most general
renormalizable interaction for a single chiral superfield. Thereby,
the auxiliary field equation of motion reads ${\cal F}^{*}+(\partial
{\cal P}/\partial {\cal A})=0$. Similarly, we can treat the vector
superfields, etc. Here we shall forbear to write them out as the
standard theory is so well known.

\section{Unaccelerated uniform motion;\,\,
a foundation of SR} Let impose peculiar constraints upon the
anticommuting spinors $(\underline{\xi},\,\underline{\bar{\xi}})$
and $(\xi,\,\bar{\xi})$:
\begin{equation}
\begin{array}{l}
\underline{\xi}^{\alpha}=i\,\frac{\underline{\tau}}{\sqrt{2}}\,\underline{\theta}^{\alpha},\quad
\underline{\bar{\xi}}_{\dot{\alpha}}=-i\,\frac{\underline{\tau}^{*}}{\sqrt{2}}\,\underline{\bar{\theta}}_{\dot{\alpha}},\quad
\xi^{\alpha}=i\frac{\tau}{\sqrt{2}} \theta^{\alpha},\quad
\bar{\xi}_{\dot{\alpha}}=-i\frac{\tau^{*}}{\sqrt{2}}
\bar{\theta}_{\dot{\alpha}},
\end{array}
\label{S83}
\end{equation}
and write down the infinitesimal displacement arisen in
$\underline{M}_{\,2}$ as
\begin{equation}
\begin{array}{l}
\Delta \eta^{\,\underline{m}}=v^{\,\underline{m}}\,\tau=
\underline{\theta}\,\sigma^{\,\underline{m}}\,\underline{\bar{\xi}}-
\underline{\xi}\,\sigma^{\,\underline{m}}\,\underline{\bar{\theta}},
\end{array}
\label{S80}
\end{equation}
where the parameter $\tau \,(=\tau^{*})$ can physically be
interpreted as the {\it duration time} of atomic double transition
of a particle from  $M_{4}$ to $\underline{M}_{\,2}$ and back. So,
\begin{equation}
\begin{array}{l}
v^{(+)}\tau=i(\underline{\theta}_{1}\,\underline{\bar{\xi}}_{1}-
\underline{\xi}_{1}\,\underline{\bar{\theta}}_{1}),\quad
v^{(-)}\tau=i(\underline{\theta}_{2}\,\underline{\bar{\xi}}_{2}-
\underline{\xi}_{2}\,\underline{\bar{\theta}}_{2}),
\end{array}
\label{S81}
\end{equation}
and that
\begin{equation}
\begin{array}{l}
 v^{2} \tau^{2}=v^{(+)}v^{(-)}\tau^{2}=
-(\underline{\theta}_{1}\,\underline{\bar{\xi}}_{1}-
\underline{\xi}_{1}\,\underline{\bar{\theta}}_{1})
(\underline{\theta}_{2}\,\underline{\bar{\xi}}_{2}-
\underline{\xi}_{2}\,\underline{\bar{\theta}}_{2})=
4\underline{\theta}_{1}\,\underline{\bar{\theta}}_{1}
\underline{\theta}_{2}\,\underline{\bar{\theta}}_{2}\geq 0.
\end{array}
\label{S82}
\end{equation}
Hance
\begin{equation}
\begin{array}{l}
v^{(+)}=
\sqrt{2}\,\underline{\theta}_{1}\,\underline{\bar{\theta}}_{1}\geq
0,\quad
v^{(-)}=\sqrt{2}\,\underline{\theta}_{2}\,\underline{\bar{\theta}}_{2}\geq
0.
\end{array}
\label{S84}
\end{equation}
According to embedding map~(\ref{S7}), therefore, we may introduce
the {\it velocity of light} $(c)$ in vacuum as maximum attainable
velocity for uniform motions of all the particles in the Minkowski
background space, $M_{4}$:
\begin{equation}
\begin{array}{l}
 c=\frac{1}{\sqrt{2}} (v^{(+)}+v^{(-)})=
\sqrt{2}\,(\underline{\theta}_{1}\,\underline{\bar{\theta}}_{1}+
\underline{\theta}_{2}\,\underline{\bar{\theta}}_{2})=
\sqrt{2}\,\underline{\theta}\,\underline{\bar{\theta}}=const,\\
 v_{q}=\frac{1}{\sqrt{2}} (v^{(+)}-v^{(-)})=
\sqrt{2}\,(\underline{\theta}_{1}\,\underline{\bar{\theta}}_{1}-
\underline{\theta}_{2}\,\underline{\bar{\theta}}_{2})=\pm|\vec{v}|=const,\quad
|\vec{v}| \leq c.
\end{array}
\label{S85}
\end{equation}
The spinors $\theta(\underline{\theta},\,\underline{\xi})$ and
$\xi(\underline{\theta},\,\underline{\xi})$ satisfy the embedding
map~(\ref{S7}), namely $\Delta q^{0}=\Delta x^{0}$ and $\Delta
q^{2}=(\Delta \vec{x})^{2}$, so from~(\ref{SS61}) we have
\begin{equation}
\begin{array}{l}
\underline{\theta}\,\sigma^{0}\,\underline{\bar{\xi}}-
\underline{\xi}\,\sigma^{0}\,\underline{\bar{\theta}}=
\theta\,\sigma^{0}\,\bar{\xi}- \xi\,\sigma^{0}\,\bar{\theta},\quad
(\underline{\theta}\,\sigma^{3}\,\underline{\bar{\xi}}-
\underline{\xi}\,\sigma^{3}\,\underline{\bar{\theta}})^{2}=
(\theta\,\vec{\sigma}\,\bar{\xi}-
\xi\,\vec{\sigma}\,\bar{\theta})^{2}.
\end{array}
\label{S79}
\end{equation}
By virtue of~(\ref{S83}), the~(\ref{S79}) reduces to
\begin{equation}
\begin{array}{l}
 \underline{\theta}_{1}\,\underline{\bar{\theta}}_{1}+
\underline{\theta}_{2}\,\underline{\bar{\theta}}_{2}=
\underline{\theta}\,\underline{\bar{\theta}}=\theta\bar{\theta},\quad
\underline{\theta}_{1}\,\underline{\bar{\theta}}_{1}-
\underline{\theta}_{2}\,\underline{\bar{\theta}}_{2}=\pm\sqrt{\frac{3}{2}}
(-\theta\theta\bar{\theta}\bar{\theta})^{1/2}=\pm\sqrt{\frac{3}{2}}\,\theta\bar{\theta},
\end{array}
\label{S86}
\end{equation}
where we use the following relations:
\begin{equation}
\begin{array}{l}

(\theta\sigma^{m}\,\bar{\theta})(\theta\sigma^{n}\,\bar{\theta})=
\frac{1}{2}\,\theta\theta\bar{\theta}\bar{\theta}\,\eta^{mn},\quad
(-\theta\theta\bar{\theta}\bar{\theta})^{1/2}=(\theta\bar{\theta}\theta\bar{\theta})^{1/2}=\theta\bar{\theta}.
\end{array}
\label{S87}
\end{equation}
So,
\begin{equation}
\begin{array}{l}
\underline{\theta}_{1}\,\underline{\bar{\theta}}_{1}=\frac{1}{2}(1
\pm\sqrt{\frac{3}{2}})\,\theta\bar{\theta},\quad
\underline{\theta}_{2}\,\underline{\bar{\theta}}_{2}=\frac{1}{2}(1
\mp\sqrt{\frac{3}{2}})\,\theta\bar{\theta}.
\end{array}
\label{S88}
\end{equation}
Hence we conclude that the {\it unaccelerated uniform motion of a
particle in $M_{4}$ is encoded in the spinors $\underline{\theta}$
and $\underline{\bar{\theta}}$ referred to the master space
$\underline{M}_{\,2}$, which is an individual companion to the
particle of interest}. Therefore, to account for the most important
two postulates constituting a foundation of SR, it would be
necessary further to impose certain constraints upon the constant
Lorentz spinors $\underline{\theta}$. Lorentz invariance is a
fundamental symmetry and refers to measurements of ideal inertial
observers that move uniformly forever on rectilinear timelike
worldlines. In view of relativity of velocity of a particle, we are
of course not limited to any particular spinor
$\underline{\theta}(\vec{v})$, but can choose at will any other
spinors $\underline{\theta}'(\vec{v'})$,
$\underline{\theta''}(\vec{v''}),\dots$ relating respectively to
velocities $\vec{v'}, \vec{v''},\dots$, whose functional dependence
(transformational law) on the original spinor
$\underline{\theta}(\vec{v})$ is known. Of the various possible
transformations, we must consider for a validity of SR only those
which obey the following constraints:
\begin{equation}
\begin{array}{l}
1.\,\,
\underline{\theta}\,\underline{\bar{\theta}}=\underline{\theta}'\,\underline{\bar{\theta}}'=
\underline{\theta}''\,\underline{\bar{\theta}}''=\cdots=\frac{c}{\sqrt{2}}=const;\\
2.\,\,
\underline{\theta}_{\,1}\,\underline{\bar{\zeta}}_{\,1}\underline{\theta}_{\,2}\,\underline{\bar{\zeta}}_{\,2}=
\underline{\theta}'_{\,1}\,\underline{\bar{\zeta}}'_{\,1}
\underline{\theta}'_{\,2}\,\underline{\bar{\zeta}}'_{\,2}=\cdots=inv.
\end{array}
\label{S89}
\end{equation}
According to first relation, we may introduce a notion of {\it
time}, for the all inertial frames of reference S, S', S",..., we
have then standard Lorentz code (SLC)-relations:\phantom{a} $x^{0}=
c t,\quad x^{0'}= c t', \quad x^{0"}= c t",\dots$. This is a second
postulate of SR (Einstein's postulate) that the velocity of light
$(c)$ in free space appears the same to all observers regardless the
relative motion of the source of light and the observer. By virtue
of second relation and equations~(\ref{S83}), we may derive the
invariant interval between the  two events defined in Minkowski
spacetime:
\begin{equation}
\begin{array}{l}
8\underline{\theta}_{\,1}\,\underline{\bar{\theta}}_{\,1}\underline{\theta}_{\,2}\,\underline{\bar{\theta}}_{\,2}\,\Delta
t^{2}= 2v^{2}\Delta t^{2}=(c^{2}-v_{q}^{2})\Delta
t^{2}=(c^{2}-\vec{v}{}^{2})\Delta t^{2}=\\ c^{2}\Delta t^{2}- \Delta
\vec{x}{}^{2}\equiv\Delta s^{2}=
8\underline{\theta}'_{\,1}\,\underline{\bar{\theta}}'_{\,1}\underline{\theta}'_{\,2}\,\underline{\bar{\theta}}'_{\,2}\,
\Delta t'{}^{2}= c^{2}\Delta t'{}^{2}- \Delta \vec{x}
'{}^{2}\equiv\Delta s'{}^{2}=\cdots=inv,
\end{array}
\label{S90}
\end{equation}
where we introduce the physical finite {\it time interval}, $\Delta
t=k\tau$, between two events as integer number of the {\it duration
time}, $\tau$, of atomic double transition of a particle from
$M_{4}$ to $\underline{M}_{\,2}$ and back, where $k$ is the number
of double transformations. Hence, an unaccelerated uniform motion,
for example, of spin-$0$ particle in $M_{4}$ can be described by the
chiral superfield
$\underline{\Phi}(\eta^{\hat{m}},\,\underline{\theta},\,\underline{\bar{\theta}})$,
while a similar motion of spin-$1/2$ particle in $M_{4}$ can be
described by the chiral superfield
$\Phi(x^{m},\,\theta,\,\bar{\theta})$, etc. So, we may refer to all
constant Lorentz spinors obeying~(\ref{S89}) as the {\it
SLC-spinors}, which  constitute a foundation of SR. Hence, in view
of the MS-SUSY mechanism of motion, the uniform motion of a particle
in $M_{4}$ is encoded in the spinors $\underline{\theta}$ and
$\underline{\bar{\theta}}$, which refer to $\underline{M}_{\,2}$.
This will call for a complete reconsideration of our ideas of
Lorentz motion code, to be now referred to as the {\it individual
code of a particle}, defined as its intrinsic property.

\subsection{Extended supersymmetry and ELC}
In four dimensions, it is possible to have as many as eight
supersymmetries: $N_{max}=4$  for renormalizable flat-space field
theories; $N_{max} = 8$ for consistent theories of supergravity. It
has been shown that the N = 4 theory is not only renormalizable but
actually finite. So, the theories with $N > 1$ may play a key role
in high-energy physics. These models explore in (\ref{S60}) more
than one distinct copy of the supersymmetry generators, $Q_{\alpha
i}$, therefore, this perspective ultimately requires to relax the
Einstein's postulate, because it is natural now to circumvent the
limitations to any particular spinor $\underline{\theta}$, instead,
considering $i(=1,\dots,4)$-th ($N_{max}=4$) copy of the spinors
$\Theta^{\alpha i}\equiv (\theta^{\alpha i}$ or
$\underline{\theta}^{\alpha i})$. Therefore, we now have a
straightforward generalization of~(\ref{S89}):
\begin{equation}
\begin{array}{l}
1.\,\,
\underline{\theta}^{i}\,\underline{\bar{\theta}}{}^{i}=\underline{\theta}^{i}{}'\,\underline{\bar{\theta}}{}^{i}{}'=
\underline{\theta}^{i}{}''\,\underline{\bar{\theta}}{}^{i}{}''=\cdots=\frac{c_{i}}{\sqrt{2}}=const;\quad
\quad\mbox{(no sum over i)}, \\
2.\,\,
\underline{\theta}_{\,1}^{i}\,\underline{\bar{\zeta}}_{\,1}^{i}\underline{\theta}_{\,2}^{i}\,
\underline{\bar{\zeta}}_{\,2}^{i}=
\underline{\theta}^{i}{}'_{\,1}\,\underline{\bar{\zeta}}^{i}{}'_{\,1}
\underline{\theta}^{i}{}'_{\,2}\,\underline{\bar{\zeta}}^{i}{}'_{\,2}=\cdots=inv.
\end{array}
\label{S99}
\end{equation}
This observation allows us to lay forth the extension of Lorentz
code, at which SLC violating new physics appears. We may now
consider the particles of  $i(=1,\dots,4)$-th {\it type}
($N_{max}=4$). That is to say, {\it the $i$-th type particle in free
Minkowski space carries an individual Lorentz motion code with its
own maximum attainable velocity $c_{i}$, i.e., its own velocity of
'light-like' state}:
\begin{equation}
\begin{array}{l}
 c_{i}=\frac{1}{\sqrt{2}} (v^{(+)}_{i}+v^{(-)}_{i})=
\sqrt{2}\,(\underline{\theta}_{1 i}\,\underline{\bar{\theta}}_{\,1
i}+ \underline{\theta}_{\,2 i}\,\underline{\bar{\theta}}_{\,2 i})=
\sqrt{2}\,\underline{\theta}_{\,i}\,\underline{\bar{\theta}}_{\,i}=const,\,
\mbox{(no sum over i)},\\
 v_{q i}=\frac{1}{\sqrt{2}} (v^{(+)}_{i}-v^{(-)}_{i})=
\sqrt{2}\,(\underline{\theta}_{\,1
i}\,\underline{\bar{\theta}}_{1\,i}- \underline{\theta}_{\, 2
i}\,\underline{\bar{\theta}}_{\,2})=\pm|\vec{v}_{i}|=const,\quad
|\vec{v}_{i}| \leq c_{i}.
\end{array}
\label{Si85}
\end{equation}
A general solution to the Lorentz covariance in the theory can be
easily accommodated if the 'time' at which event occurs is extended
by allowing an extra dependence on 'different type' readings $t_{i}$
referred to the particles of different type. They satisfy for all
inertial frames of reference S, S', S",..., so-called
'ELC-relations':
\begin{equation}
\begin{array}{l}
x^{0}\equiv c_{1}t_{1}=\dots=c_{i}t_{i}=\dots,\\
x^{0'}\equiv c_{1}t'_{1}=\dots=c_{i}t'_{i}=\dots,\\
\dots\dots\dots\dots\dots\dots\dots\dots\dots\dots
\end{array}
\label{S100}
\end{equation}
where $c_{1}\equiv c$ is the speed of light in vacuum, and $c_{i}>
c_{1}, \,(i=2,3,4)$ are speeds of the additional 'light-like'
states, higher than that of light. The clock reading $t_{i}$ can be
used for the $i-$th type particle, the velocity of which reads
$v_{i}=x/t_{i}=c_{i}x/x^{0}$, so $\beta=v_{1}/c_{1}=\dots
v_{i}/c_{i}=\dots\equiv v/c= x/x^{0}$. If $v_{i}=c_{i}$ then
$v_{1}=c_{1}$, and the proper time of 'light-like' states are
described by the null vectors $d s^{2}_{1}=\dots d
s^{2}_{i}=\dots=0$.  The extended Lorentz transformation equations
for given $i$-th and $j$-th type clock readings can be written then
in the form
\begin{equation}
\begin{array}{l}
 x'=\gamma (x-vt),\quad t'_{i}=\gamma
\frac{c_{j}}{c_{i}}(t_{j}-\frac{v_{j}}{c_{j}^{2}}x), \quad
y'=y,\quad z'=z,\quad \gamma\equiv 1/\sqrt{1-\beta^{2}}.
\end{array}
\label{S101}
\end{equation}
Hence, like the standard SR theory, regardless the type of clock, a
metre stick traveling with system S measures shorter in the same
ratio, when the simultaneous positions of its ends are observed in
the other system S': $dx'=dx/\gamma$. Furthermore, a time interval
$dt_{i}$ specified by the $i-$th type readings, which occur at the
same point in system S ($dx=0$), will be specified with the $j-$th
type readings of system S' as $dt'_{j}=\gamma(c_{i}/c_{j})dt_{i}$.
Here we have called attention to the fact that the mere composition
of velocities which are not themselves greater than that of $c_{i}$
will never lead to a speed that is greater than that of $c_{i}$.
Inevitably in the ELC-framework a specific task is arisen then to
distinguish the type of particles. This evidently cannot be done
when the velocity ranges of  different type particles intersect. To
reconcile this situation, we note that, according to (\ref{S100}),
we may freely interchange the types of particles in the
intersection. Therefore, we adopt following convention. With no loss
of generality, we may re-arrange a general solution that the
particles with velocities $v_{1}< c_{1}$, regardless their type,
will be treated as the 1-th type particles and, thus,  a common
clock reading for them and light will be set as $t_{1}$. This part
of a formalism  is completely equivalent to the SLC-framework.
Successively, the particles, other than 'light-like' ones, with
velocities in the range $c_{i-1}\leq v_{i}< c_{i}$, regardless their
type, will be treated as the i-th type particles and, thus, a common
clock reading for them and 'light-like' state $(i)$ will be set as
$t_{i}$. The invariant momentum
\begin{equation}
\begin{array}{l}
 p^{2}_{i}=p_{\mu
i}p^{\mu}_{i}=\left(\frac{E_{i}}{c_{i}}\right)^{2}-\vec{p}_{i}^{2}=m_{0\,i}^{2}c_{i}^{2}=
p^{2}_{1}=p_{\mu
1}p^{\mu}_{1}=\left(\frac{E_{1}}{c_{1}}\right)^{2}-\vec{p}_{1}^{2}=m_{0}^{2}c_{1}^{2},
\end{array}
\label{S103}
\end{equation}
introduces a {\it modified dispersion relation} for $i-$th type
particle:
\begin{equation}
\begin{array}{l}
E_{i}^{2}=\vec{p}_{i}^{2}c_{i}^{2}+m_{0
i}^{2}c_{i}^{4}=\vec{p}_{i}^{2}c_{i}^{2}+m_{0
1}^{2}c_{1}^{2}c_{i}^{2},
\end{array}
\label{D103}
\end{equation}
where the mass of $i-$th type particle has the value $m_{0\,i}$,
when at rest,  the positive energy is
\begin{equation}
E_{i}=m_{i}c_{i}^{2}=\gamma m_{0 i}c_{i}^{2}=\gamma m_{0
1}c_{1}c_{i}, \label{S104}
\end{equation}
and $\vec{p}_{i}=m_{i}\vec{v}_{i}=\gamma m_{0 i}\vec{v}_{i}$ is the
momentum. The relation~(\ref{S104}) modifies the well-known
Einstein's equation that energy $E$ always has immediately
associated with it a positive mass $m_{i}=\gamma m_{0 i}$, when
moving with the velocity $\vec{v}_{i}$. Having set this theoretical
background, one may find  some consequences for the superluminal
propagation of particles. In particular, in the ELC-framework of
uniform motion, the time elapsing between the cause $t_{iA}$ and its
effect $t_{iB}$ as measured  for the $i-$th type superluminal
particle is
\begin{equation}
\begin{array}{l}
\Delta t_{i}=t_{iB}-t_{iA}=\frac{x_{B}-x_{A}}{v_{i}},
\end{array}
\label{S110}
\end{equation}
where $x_{A}$ and $x_{B}$ are the coordinates of the two points A
and B. In another system S', which is chosen as before and has the
arbitrary velocity $V\equiv V_{j}$ with respect to S, the time
elapsing between cause and effect would be
\begin{equation}
\begin{array}{l}
\Delta
t'_{i}=\frac{1-\frac{V_{j}}{c_{j}}\frac{v_{i}}{c_{i}}}{\sqrt{1-\frac{V^{2}_{j}}{c_{j}^{2}}}}\Delta
t_{i}\geq 0,
\end{array}
\label{S111}
\end{equation}
where according to ~(\ref{S99}), $t_{iB}=(c_{j}/c_{i})t_{jB}$ and
$t_{iA}=(c_{j}/c_{i})t_{jA}$. That is, {\it the ELC-framework
recovers the causality for a superluminal propagation}, so the
starting of the superluminal impulse at A and the resulting
phenomenon at B are being connected by the relation of cause and
effect in arbitrary inertial frames. Furthermore, in this framework,
we may give a justification of forbiddance of Vavilov-Cherenkov
radiation/or analog processes in vacuum. Thereby, in this framework
we have to set, for example, $k_{1}=(\frac{\omega}{ c_{1}},
\vec{k_{1}})$ for the 1-th type $\gamma_{1}$ photon, provided
$\vec{k_{1}}=\vec{e}_{k}\frac{\omega}{c_{1}}$, and
$p_{2}=(\frac{E_{2}}{ c_{2}}, \vec{p}_{2})$ for the 2-nd type
superluminal particle. Then the process ($l_{2}\rightarrow
l_{2}+\gamma_{1}$) becomes kinematically permitted if and only if
\begin{equation}
\begin{array}{l} k_{1} p_{2}=\frac{\omega}{c_{1}}
\frac{E_{2}}{
c_{2}}\left(1-\vec{e}_{k}\frac{\vec{v}_{2}}{c_{2}}\right)= 0,
\end{array}
\label{S113}
\end{equation}
which yields  $\omega \equiv 0$ because of
$\left(1-\vec{e}_{k}\frac{\vec{v}_{\nu 2}}{c_{2}}\right)\neq 0$.
This evades constraints due to VC-like processes since the
superluminal particle $\nu_{\mu 2}$ does not actually travel faster
than the speed $c_{2}$. Finally, in ELC-framework we discuss the
VC-radiation  of the charged superluminal particle propagating in
vacuum with a constant speed $v_{2}> c_{1}$ higher than that of
light. Recall that, for a charged particle ($e\neq 0$) moving in a
transparent, isotropic and non-magnetic medium with a constant
velocity higher than velocity of light in this medium the VC
radiation is allowed. The energy loss per frequency is~\cite{La}
\begin{equation}
\begin{array}{l}
d F= -d\omega \frac{ie^{2}}{2\pi}\sum
\omega\left(\frac{1}{c^{2}}-\frac{1}{\varepsilon
v^{2}}\right)\int\frac{d\zeta}{\zeta},
\end{array}
\label{R25}
\end{equation}
where the direction of the velocity $\vec{v}$ is chosen to be
$x-$direction: $k_{x}=k\cos \theta=\omega/v$, $\,k=n\omega/c$ is the
wave number $n=\sqrt{\varepsilon}$ is the real refractive index,
$\varepsilon$ is the permittivity. The summation is over terms with
$\omega=\pm|\omega|$, and a variable
\begin{equation}
\begin{array}{l}
\zeta=q^{2}-\omega^{2}\left(\frac{\varepsilon}{c^{2}}-\frac{1}{v^{2}}\right)
\end{array}
\label{R26}
\end{equation}
is introduced, provided $q=\sqrt{k_{y}^{2}+k_{z}^{2}}$. The
integrand in~(\ref{R25}) is strongly peaked near the singular point
$\zeta=0$, for which $q^{2}+k_{x}^{2}=k^{2}$.  Using standard
technique, the  formula (\ref{R25}) can  be easily further
transformed to be applicable in ELC-framework for the charged
superluminal particle of $2$-nd type propagating in vacuum (i.e. if
$\varepsilon=1$) with a constant speed $v_{2}$ higher than that of
light ($c_{1}\leq v_{2}< c_{2})$:
\begin{equation}
\begin{array}{l}
d F= -d\omega \frac{ie^{2}}{2\pi}\sum
\omega\left(\frac{1}{c^{2}_{2}}-\frac{1}{v^{2}_{2}}\right)\int\frac{d\zeta}{\zeta},
\end{array}
\label{R27}
\end{equation}
and, respectively, (\ref{R26}) becomes
\begin{equation}
\begin{array}{l}
\zeta=q^{2}_{1}-\omega^{2}\left(\frac{1}{c^{2}_{2}}-\frac{1}{v^{2}_{2}}\right),
\end{array}
\label{R28}
\end{equation}
where $q_{1}=\sqrt{k_{y1}^{2}+k_{z1}^{2}}$, $\,q^{2}_{1}+k_{x1}^{2}=
k_{1}^{2}=\omega^{2}/c_{1}^{2}$, and now $k_{x1}v_{2}=\omega$. We
have then
\begin{equation}
\begin{array}{l}
\zeta=\frac{\omega^{2}}{c^{2}_{2}}\left(
\frac{c_{2}^{2}}{v^{2}_{2}\cos^{2}\theta}-1\right)\neq 0,
\end{array}
\label{R29}
\end{equation}
because of $v_{2}< c_{2}$, and that the integral~(\ref{R27}) is
zero, since the integrand has no poles. Hence, as expected, the {\it
VC-radiation of a charged superluminal  particle as it propagates in
vacuum is forbidden}.

\section{Accelerated motion and local MS-SUSY} In case of
an accelerated $(a=|\vec{a}|\neq 0)$ motion of a particle in
$M_{4}$, according to~(\ref{SS61}), we have then
\begin{equation}
\begin{array}{l}
\frac{i}{\sqrt{2}}\left(\underline{\theta}\,\sigma^{3}\frac{d^{2}\bar{\xi}}{dt^{2}}-
\frac{d^{2}\xi}{dt^{2}}\sigma^{3}\underline{\bar{\theta}}\right)=\frac{d^{2}q}{d
t^{2}}=a=\frac{1}{\sqrt{2}}\left(\frac{d^{2} \eta^{(+)}}{d
t^{2}}-\frac{d^{2} \eta^{(-)}}{d t^{2}}\right)=
\frac{1}{\sqrt{2}}\left(a^{(+)}-a^{(-)}\right),\quad
a^{(\pm)}=\frac{d v^{(\pm)}}{d t}.
\end{array}
\label{T1}
\end{equation}
So, we may relax the condition $\partial_{\hat{m}} \epsilon=0$ and
promote this symmetry to a local supersymmetry in which the
parameter $\epsilon=\epsilon(X^{\hat{m}})$ depends explicitly on
$X^{\hat{m}}$. Such a local SUSY can already be read off from the
algebra~(\ref{S47}) in the form
\begin{equation}
\begin{array}{l}
[\epsilon(X) Q,\,
\bar{Q}\bar{\epsilon}(X)]=2\epsilon(X)\sigma^{\hat{m}}\bar{\epsilon}(X)p_{\hat{m}},
\end{array}
\label{T3}
\end{equation}
which says that the product of two supersymmetry transformations
corresponds to a translation in space-time of which the four
momentum $p_{\hat{m}}$ is the generator. Similar to the results of
subsection F, the multiplication of two local successive
transformations induces the motion
\begin{equation}
\begin{array}{l}
g(0,\,\epsilon(X),\,\bar{\epsilon}(X))\,\Omega
(X^{\hat{m}},\,\Theta,\,\bar{\Theta})\,\longrightarrow\,
(X^{\hat{m}}+i\,\Theta\,\sigma^{\hat{m}}\,\bar{\epsilon}(X)-i\,\epsilon(X)\,\sigma^{\hat{m}}\,\bar{\Theta},\,
\Theta+\epsilon(X),\,\bar{\Theta}+\bar{\epsilon}(X)),
\end{array}
\label{T4}
\end{equation}
and, in accord, the transformation~(\ref{S51}) is expected to be
somewhat of the form
\begin{equation}
\begin{array}{l}
[\delta_{\epsilon_{1}(X)},\,
\delta_{\epsilon_{2}(X)}]V\sim\epsilon_{1}(X)\sigma^{\hat{m}}\bar{\epsilon}_{2}(X)\,\partial_{\hat{m}}V,
\end{array}
\label{T5}
\end{equation}
that differ from point to point, namely this is the notion of a
general coordinate transformation. Whereupon we see that for the
local MS-SUSY to exist it requires the background spaces
$(\widetilde{\underline{M}_{\,2}},\, \widetilde{M}_{4})$ to be
curved.  Thereby, the space $\widetilde{M}_{4}$, in order to become
on the same footing with the distorted space
$\underline{\widetilde{M}}_{\,2}$, refers to the accelerated proper
reference frame of a particle, without relation to other matter
fields. A useful guide in the construction of local superspace is
that it should admit rigid superspace as a limit. The reverse is
also expected, since if one starts with a constant parameter $
\epsilon$~(\ref{S46})  and performs a local Lorentz transformation,
then this parameter will in general become space-time dependent as a
result of this Lorentz transformation. The mathematical structure of
the local MS-SUSY theory has much in common with those used in the
geometrical framework of standard supergravity theories. In its
simplest version, supergravity was conceived  as a quantum field
theory whose action included the Einstein-Hilbert term, where the
graviton coexists with a fermionic field called gravitino, described
by the Rarita-Scwinger kinetic term. The two fields differ in their
spin: $2$ for the graviton, $3/2$ for the gravitino. The different
4D $\,N=1$ supergravity multiplets  all contain the graviton and the
gravitino, but differ by their systems of auxiliary fields. For a
detailed discussion we refer to the papers by ~\cite{FF}-\cite{WB},
\cite{Nil}, \cite{We}, \cite{Ja}, \cite{Bi}, \cite{Din}-\cite{DZ}.
These fields would transform into each other under local
supersymmetry. We may use the usual language which is almost
identical to the vierbien formulation of GR with some additional
input. In this framework supersymmetry and general coordinate
transformations are described in a unified way as certain
diffeomorphisms. The motion~(\ref{T4}) generates the super-general
coordinate reparametrization
\begin{equation}
\begin{array}{l}
z^{M}\,\longrightarrow\,z'{}^{M}=z^{M}-\zeta^{M}(z),
\end{array}
\label{T6}
\end{equation}
where $\zeta^{M}(z)$ arc arbitrary functions of $z$. The dynamical
variables of superspace formulation are the frame field $E^{A}(z)$
and connection $\Omega$. The superspace
$(z^{M},\,\Theta,\,\bar{\Theta})$ has at each point a tangent
superspace spanned by the frame field
$E^{A}(z)=dz^{M}E^{\phantom{a}A}_{M}(z)$, defined as a 1-form over
superspace, with coefficient superfields, generalizing the usual
frame, namely supervierbien $E^{\phantom{a}A}_{M}(z)$. Here, we use
the first half of capital Latin alphabet $A, B, \dots $ to denote
the anholonomic indices related to the tangent superspace structure
group, which is taken to be just the Lorentz group. The formulation
of supergravity in superspace provides a unified description of the
vierbein and the Rarita-Schwinger fields. They are identified in a
common geometric object, the local frame $E^{A}(z)$ of superspace.
Covariant derivatives with respect to local Lorentz transformations
are constructed by means of the spin connection, which is a 1-form
in superspace as well. Here we shall forbear to write the details
out as the standard theory is so well known. The super-vierbien
$E_{M}^{\phantom{A}A}$ and spin- connection $\Omega$ contain many
degrees of freedom. Although some of these are removed by the
tangent space and supergeneral coordinate transformations, there
still remain many degrees of freedom. There is no general
prescription for deducing necessary covariant constraints which if
imposed upon the superfields of super-vierbien and spin-connection
will eliminate the component fields. However, some usual constraints
can be found using tangent space and supergeneral coordinate
transformations of the torsion and curvature covariant tensors,
given in appropriate super-gauge. Together with other details of the
theory, they can be seen in the textbooks, see e.g.~\cite{WB},
\cite{We}. The final form of transformed super-vierbien, can be
written as
\begin{equation}
\begin{array}{l}
\left.E^{\phantom{a} M}_{A}(z)\right|_{\Theta=\bar{\Theta}=0}=\left(
                        \begin{array}{cll}
                          e^{\phantom{a} \hat{a}}_{\hat{m}}(X) &\,
                          \frac{1}{2}\psi_{\hat{m}}^{\phantom{a}
                          \alpha}(X)
                          &\,\frac{1}{2}\bar{\psi}_{\hat{m}\dot{\alpha}}(X) \\\\
                          0
                          &\,
                          \delta^{\mu}_{\alpha} &\, 0 \\
                          0
                          &\, 0 &\, \delta^{\mu}_{\dot{\alpha}} \\
                        \end{array}
                      \right),

\end{array}
\label{T23}
\end{equation}
where the fields of graviton $e^{\phantom{a} \hat{a}}_{\hat{m}}$ and
gravitino $\frac{1}{2}\psi_{\hat{m}}^{\phantom{a}
                          \alpha},\,\frac{1}{2}\bar{\psi}_{\hat{m}\dot{\alpha}}$
cannot be gauged away. Provided, we have
\begin{equation}
\begin{array}{l}
e^{\phantom{a} \hat{m}}_{\hat{a}}e^{\phantom{a}
\hat{b}}_{\hat{m}}=\delta^{\hat{b}}_{\hat{a}},\quad
\psi_{\hat{a}}^{\phantom{a} \mu}=e^{\phantom{a}
\hat{m}}_{\hat{a}}\psi_{\hat{m}}^{\phantom{a}
                          \alpha}\delta^{\mu}_{\alpha} ,\quad
\bar{\psi}_{\hat{a}\dot{\mu}}=e^{\phantom{a}
\hat{m}}_{\hat{a}}\bar{\psi}_{\hat{m}\dot{\alpha}}\delta_{\mu}^{\dot{\alpha}}.
\end{array}
\label{T24}
\end{equation}
The tetrad field $e^{\phantom{a} \hat{a}}_{\hat{m}}(X)$ plays the
role of a gauge field associated with local transformations. The
Majorana type field $\frac{1}{2}\psi_{\hat{m}}^{\phantom{a} \alpha}$
is the gauge field related to local supersymmetry. These two fields
belong to the same supergravity multiplet which also accommodates
auxiliary fields so that the local supersymmetry algebra closes.
Under infinitesimal transformations of local supersymmetry, they
transformed as
\begin{equation}
\begin{array}{l}
\delta e^{\phantom{a} \hat{a}}_{\hat{m}}=i(\psi_{\hat{m}}
\sigma^{\hat{a}}\zeta-\zeta\sigma^{\hat{a}}\bar{\psi}_{\hat{m}}),\\
\delta \psi_{\hat{m}}=-2{\cal
D}_{\hat{m}}\zeta^{\alpha}+ie^{\phantom{a} \hat{c}}_{\hat{m}}\{
\frac{1}{3}M(\varepsilon
\sigma_{\hat{c}}\bar{\zeta})^{\alpha}+b_{\hat{c}}\zeta^{\alpha}+\frac{1}{3}b^{\hat{d}}(\zeta\sigma_{\hat{d}}\bar{\sigma}_{\hat{c}})\},
\end{array}
\label{T25}
\end{equation}
etc., where $M_{4}$ and $b_{\bar{a}}$ are the auxiliary fields, and
$\zeta^{\alpha}(z)=\zeta^{\alpha}(X), \quad
\bar{\zeta}^{\alpha}(z)=\bar{\zeta}^{\alpha}(X)$ and
$\zeta^{\bar{a}}(z)=2i[\Theta\sigma^{\hat{a}}\bar{\zeta}(X)-\zeta(X)\sigma^{\hat{a}}\bar{\Theta}]$.
The chiral superfields are defined as $ \bar{{\cal
D}}_{\dot{\alpha}}\Phi=0, $ therefore, the components fields are
\begin{equation}
\begin{array}{l}
{\cal A}=\left.\Phi\right|_{\Theta=\bar{\Theta}=0},\quad
\psi_{\alpha}=\frac{1}{\sqrt{2}}{\cal
D}_{\alpha}\left.\Phi\right|_{\Theta=\bar{\Theta}=0},\quad {\cal
F}=-\frac{1}{4}{\cal D}^{\alpha}{\cal
D}_{\alpha}\left.\Phi\right|_{\Theta=\bar{\Theta}=0},
\end{array}
\label{T27}
\end{equation}
which carry Lorentz indices. Under infinitesimal transformations of
local supersymmetry, they transformed as
\begin{equation}
\begin{array}{l}
\delta{\cal A}=-\sqrt{2}\,\zeta^{\alpha}\psi_{\alpha},\\
\delta\psi_{\alpha}=-\sqrt{2}\,\zeta_{\alpha}{\cal
F}-i\sqrt{2}\,\sigma_{\alpha\dot{\beta}}^{\phantom{ab}\hat{a}}\bar{\zeta}{}^{\dot{\beta}}
\hat{\cal D}_{\hat{a}}{\cal A},\\
\delta{\cal
F}=-\frac{1}{3}\sqrt{2}\,M^{*}\zeta^{\alpha}\psi_{\alpha}+\bar{\zeta}^{\dot{\alpha}}(\frac{1}{6}
\sqrt{2}\,b_{\alpha\dot{\alpha}}\psi^{\dot{\alpha}}-i\sqrt{2}\,\hat{\cal
D}_{\alpha\dot{\alpha}}\psi^{\alpha}),
\end{array}
\label{T28}
\end{equation}
where $\hat{\cal D}_{\hat{a}}$ is, so-called, super-covariant
derivative
\begin{equation}
\begin{array}{l}
\hat{\cal D}_{\hat{a}}{\cal A}\equiv e^{\phantom{a}
\hat{m}}_{\hat{a}}(\partial_{\hat{m}}{\cal
A}-\frac{i}{\sqrt{2}}\psi^{\phantom{a} \mu}_{\hat{m}}\psi_{\mu}),\\
\hat{\cal D}_{\hat{a}}\psi_{\alpha}=e^{\phantom{a}
\hat{m}}_{\hat{a}}({\cal
D}_{\hat{m}}\psi_{\alpha}-\frac{1}{\sqrt{2}}\psi_{\hat{m}\alpha}{\cal
F}-\frac{i}{\sqrt{2}}\bar{\psi}_{\hat{m}}^{\phantom{a}
\dot{\beta}}\hat{\cal D}_{\alpha\dot{\beta}}{\cal A}).
\end{array}
\label{T29}
\end{equation}
The graviton and the gravitino form thus the basic multiplet of
local MS-SUSY, and one expects the simplest locally supersymmetric
model to contain just this multiplet. The spin $3/2$ contact term in
total Lagrangian arises from equations of motion for the torsion
tensor, and that the original Lagrangian itself takes the simpler
interpretation of a minimally coupled spin $(2,\,3/2)$ theory.

\section{Inertial effects}
We would like to place the emphasis on the essential difference
arisen between the standard supergravity theories and some rather
unusual properties of local MS-SUSY theory. In the framework of the
standard supergravity theories, as in GR, a curvature of the
space-time acts on all the matter fields. The source of graviton is
the energy-momentum tensor of matter fields, while the source of
gravitino is the spin-vector current of supergravity. In the local
MS-SUSY theory, unlike the supergravity, a curvature of space-time
arises entirely due to the inertial properties of the
Lorentz-rotated frame of interest, i.e. a {\it fictitious
gravitation} which can be globally removed by appropriate coordinate
transformations. This refers to the particle of interest itself,
without relation to other matter fields. The only source of graviton
and gravitino, therefore, is the acceleration of a particle, because
the MS-SUSY is so constructed as to make these two particles just as
being the two bosonic and fermionic states of a particle of interest
in the curved background spaces $\widetilde{M}_{4}$ and
$\widetilde{\underline{M}}_{\,2}$, respectively, or vice versa.
Whereas,  in order to become on the same footing with the distorted
space $\underline{\widetilde{M}}_{\,2}$, the space
$\widetilde{M}_{4}$ refers only to the accelerated proper reference
frame of a particle. With these physical requirements, a standard
Lagrangian consisted of the classical Einstein-Hilbert Lagrangian
plus a part which contains the Rarita-Schwinger field and coupling
of supergravity with matter superfields evidently no longer holds.
Instead we are now looking for an alternative way of implications of
local MS-SUSY in the model of accelerated motion and inertial
effects. For example, we may with equal justice start from the
reverse, which as we mentioned before is also expected. If one
starts with a constant parameter $ \epsilon$~(\ref{S46})  and
performs a local Lorentz transformation, which can only be
implemented if MS and space-time are curved (deformed/distorted) $
(\widetilde{M}_{2},\,\widetilde{M}_{4})$, then this parameter will
in general become space-time dependent as a result of this Lorentz
transformation, which readily implies {\it local} MS-SUSY. In going
into practical details of the realistic local MS-SUSY model, it
remains to derive the explicit form of the vierbien $ e^{\phantom{a}
\hat{a}}_{\hat{m}}(\varrho)\equiv(e^{\phantom{a}
a}_{m}(\varrho),\,e^{\phantom{a}
\underline{a}}_{\underline{m}}(\varrho)), $ which describes {\it
fictitious graviton} as a function of {\em local rate}
$\varrho(\eta,m,f)$ of instantaneously change of the velocity
$v^{(\pm)}$ of massive $(m)$ test particle under the unbalanced net
force $(f)$. At present, unfortunately, we cannot offer a
straightforward recipe for deducing the alluded vierbien $
e^{\phantom{a} \hat{a}}_{\hat{m}}(\varrho)$ in the framework of
quantum field theory of MS-supergravity. However, recently it was
derived by~\cite{gago1} in the framework of classical physics.
Together with other usual aspects of the theory, this illustrates a
possible solution to the problems of inertia behind spacetime
deformations. Thereby it was argued that a deformation/(distortion
of local internal properties)  of $\underline{M}_{\,2}$ is the
origin of inertia effects that can be observed by us. Consequently,
the next member of the basic multiplet of local MS-SUSY -{\it
fictitious gravitino},
$\psi_{\hat{m}}^{\phantom{a}\alpha}(\varrho)$, will be arisen under
infinitesimal transformations of local supersymmetry~(\ref{T25}),
provided by the local parameters $\zeta^{M}(a)$~(\ref{T6}).

\subsection{The general space-time deformation/distortion of MS}
For the self-contained arguments we first review of certain
essential theoretical aspects of a general {\em distortion of local
internal properties of MS}~\cite{gago1}, formulated in the framework
of classical physics. There was no need for a major revision of the
topic but we have taken the opportunity to make one improvement.
That is, we show how this recovers the {\em world-deformation
tensor} $\widetilde{\Omega}$, which still has to be put in
\cite{gago2} by hand.  To start with, let $\underline{V}_{\,2}$ be
2D semi-Riemann space, which has at each point a tangent space,
$\breve{T}_{\breve{\eta}}\underline{V}_{\,2}$, spanned by the
anholonomic orthonormal frame field, $\breve{e}$, as a shorthand for
the collection of the 2-tuplet
$(\breve{e}_{(+)},\,\breve{e}_{(-)})$, where
$\breve{e}_{\hat{a}}=\breve{e}_{\hat{a}}^{\phantom{a}\underline{\mu}}\,
\breve{e}_{\underline{\mu}}$, with the holonomic frame is given as
$\breve{e}_{\underline{\mu}}= \breve{\partial}_{\underline{\mu}}$.
Here, we use the first half of Latin alphabet $\hat{a}, \hat{b},
\hat{c}, . . . = (\pm)$ to denote the anholonomic indices related to
the tangent space, and the letters
$\underline{\mu},\underline{\nu}=\tilde{(\pm)}$ to denote the
holonomic world indices related  either to the space
$\underline{V}_{\,2}$ or $\underline{\widetilde{{M}}}_{2}$. All
magnitudes referred to the space, $\underline{V}_{\,2}$, will be
denoted with an over
${}^{\prime}\phantom{a}\breve{}\phantom{a}{}^{\prime}$. These then
define a dual vector, $\breve{\vartheta}$, of differential forms, $
\breve{\vartheta}=\left(
                    \begin{array}{c}
                      \breve{\vartheta}{}^{(+)} \\
                      \breve{\vartheta}{}^{(-)} \\
                    \end{array}
                  \right),
$ as a shorthand for the collection of the
$\breve{\vartheta}{}^{\hat{b}} =\breve{e}{}^{\hat{b}}_{\phantom{a}
\underline{\mu}}\,\breve{\vartheta}{}^{\underline{\mu}}$, whose
values at every point form the dual basis, such that
$\breve{e}_{\hat{a}}\,\rfloor\,
\breve{\vartheta}{}^{\hat{b}}=\delta^{\hat{b}}_{\hat{a}}$, where
$\rfloor$ denoting the interior product, namely, this is a
$C^{\infty}$-bilinear map $\rfloor:\Omega^{1}\rightarrow \Omega^{0}$
with $\Omega^{p}$ denotes the $C^{\infty}$-modulo of differential
p-forms on $\underline{V}_{\,2}$. In components
$\breve{e}_{\hat{a}}^{\phantom{a}\underline{\mu}}\,\breve{e}{}^{\hat{b}}_{\phantom{a}\underline{\mu}}=\delta^{\hat{b}}_{\hat{a}}$.
On the manifold, $\underline{V}_{\,2}$, the tautological tensor
field, $i\breve{d}$, of type (1,1) can be defined which assigns to
each tangent space the identity linear transformation. Thus for any
point $\breve{\eta}\in \underline{V}_{\,2}$, and any vector
$\breve{\xi}\in \breve{T}_{\breve{\eta}}\underline{V}_{\,2}$, one
has $i\breve{d}(\breve{\xi})=\breve{\xi}$. In terms of the frame
field, the $\breve{\vartheta}{}^{\hat{a}}$ give the expression for
$i\breve{d}$ as
$i\breve{d}=\breve{e}\breve{\vartheta}=\breve{e}_{(+)}\otimes\breve{\vartheta}{}^{(+)}+
\breve{e}_{(-)}\otimes\breve{\vartheta}{}^{(-)}$, in the sense that
both sides yield $\breve{\xi}$ when applied to any tangent vector
$\breve{\xi}$ in the domain of definition of the frame field. We may
consider general transformations of the linear group, $GL(2, R)$,
taking any base into any other set of four linearly independent
fields. The notation,
$\{\breve{e}_{\hat{a}},\,\breve{\vartheta}{}^{\hat{b}}\}$, will be
used below for general linear frames. The holonomic metric can be
defined in the semi-Riemann space, $\underline{V}_{\,2}$, as
\begin{equation}
\begin{array}{l}
\breve{g}=\breve{g}_{\underline{\mu}\underline{\nu}}\,\breve{\vartheta}{}^{\underline{\mu}}\otimes\breve{\vartheta}{}^{\underline{\nu}}=
\breve{g}(\breve{e}_{\underline{\mu}},
\,\breve{e}_{\underline{\nu}})\,
\breve{\vartheta}{}^{\underline{\mu}}\otimes\breve{\vartheta}{}^{\underline{\nu}},
\end{array}
\label{R24}
\end{equation}
with components,
$\breve{g}_{\underline{\mu}\underline{\nu}}=\breve{g}(\breve{e}_{\underline{\mu}},
\breve{e}_{\underline{\nu}})$ in the dual holonomic base
$\{\breve{\vartheta}{}^{\underline{\mu}}\}$. The anholonomic
orthonormal frame field, $\breve{e}$, relates $\breve{g}$ to the
tangent space metric, ${}^{*}o_{\hat{a}\hat{b}}$, by $
{}^{*}o_{\hat{a}\hat{b}} = \breve{g}(\breve{e}_{\hat{a}},
\,\breve{e}_{\hat{b}})=
\breve{g}_{\underline{\mu}\underline{\nu}}\,\breve{e}_{\hat{a}}^{\phantom{a}\underline{\mu}}\,
\breve{e}_{\hat{b}}^{\phantom{a}\underline{\nu}} $, which has the
converse
$\breve{g}_{\underline{\mu}\underline{\nu}}={}^{*}o_{\hat{a}\hat{b}}\,\breve{e}{}^{\hat{a}}_{\phantom{a}\underline{\mu}}\,
\breve{e}{}^{\hat{b}}_{\phantom{a}\underline{\nu}}$ because of the
relation
$\breve{e}_{\hat{a}}^{\phantom{a}\underline{\mu}}\,\breve{e}{}^{\hat{a}}_{\phantom{a}\underline{\nu}}=
\delta^{\underline{\mu}}_{\underline{\nu}}$.  A {\em distortion of
local internal
properties of MS} comprises then two steps.\\
1) The linear frame
$(\underline{e}_{\,\underline{m}};\,\underline{\vartheta}{}^{\,\underline{m}})$
at given point ($p\in \underline{M}_{\,2}$) is undergone the {\em
distortion} transformations conducted by  $(\breve{D},\,\breve{Y})$
and $(D,\, Y)$,  relating respectively to $\underline{V}_{\,2}$ and
$\underline{\widetilde{{M}}}_{2}$, ,\ which may be recast in the
form
\begin{equation}
\begin{array}{l}
\breve{e}_{\underline{\mu}}=\breve{D}{}_{\underline{\mu}}^{\underline{m}}\,\bar{e}_{\underline{m}},\quad
\breve{\vartheta}{}^{\underline{\mu}}=\breve{Y}{}^{\underline{\mu}}_{\underline{m}}\,\bar{\vartheta}{}^{\underline{m}},\quad
e_{\underline{\mu}}=D_{\underline{\mu}}^{\underline{m}}\,\bar{e}_{\underline{m}},\quad
\vartheta^{\underline{\mu}}=Y^{\underline{\mu}}_{\underline{m}}\,\bar{\vartheta}{}^{\underline{m}}.
\end{array}
\label{D1}
\end{equation}
2) The norm $d\widetilde{\hat\eta}\equiv id$ of the infinitesimal
displacement $d\widetilde{\eta}{}^{\tilde{A}}$ on the general smooth
differential 2D-manifold $\widetilde{\mathcal{M}}_{2}$ can then be
written in terms of the space-time structures of $V_{2}$ and $M_{2}$
as
\begin{equation}
\begin{array}{l}

id=e\,\vartheta=\widetilde{\Omega}{}_{\underline{\mu}}^{\phantom{a}\underline{\nu}}\,\breve{e}_{\underline{\nu}}\otimes
\breve{\vartheta}{}^{\underline{\mu}}=\Omega_{\hat{b}}^{\phantom{a}\hat{a}}\,\breve{e}_{\hat{a}}\otimes\breve{\vartheta}{}^{\hat{b}}=
e_{\underline{\mu}}\otimes\vartheta^{\underline{\mu}}=e_{\hat{a}}\otimes\vartheta^{\hat{a}}=
\Omega^{\phantom{a}\underline{n}}_{\underline{m}}\,\bar{e}_{\underline{n}}\otimes\bar{\vartheta}{}^{\underline{m}}
\,\in\,\underline{\widetilde{M}}_{\,2},
\end{array}
\label{D3}
\end{equation}
where
$e=\{e_{\hat{a}}=e_{\hat{a}}^{\phantom{a}\underline{\mu}}\,e_{\underline{\mu}}\}$
is the frame field and
$\vartheta=\{\vartheta^{\hat{a}}=e^{\hat{a}}_{\phantom{a}\underline{\mu}}\,\vartheta^{\underline{\mu}}\}$
is the coframe field defined on $\underline{\widetilde{M}}_{\,2}$,
such that $e_{\hat{a}}\,\rfloor\,
\vartheta^{\hat{b}}=\delta^{\hat{b}}_{\hat{a}}$. Hence the
anholonomic deformation tensor
$\Omega^{\phantom{a}\hat{a}}_{\hat{b}}=\pi^{\phantom{a}\hat{a}}_{\hat{c}}\,\pi^{\phantom{a}\hat{c}}_{\hat{b}}=
\widetilde{\Omega}{}^{\phantom{a}\underline{\nu}}_{\underline{\mu}}\,\breve{e}{}^{\hat{a}}_{\phantom{a}\underline{\nu}}
\,\breve{e}_{\hat{b}}^{\phantom{a}\underline{\mu}}\,\,$ yields local
tetrad deformations
\begin{equation}
\begin{array}{l}
e_{\hat{c}}=\pi^{\phantom{a}\hat{a}}_{\hat{c}}\,\breve{e}_{\hat{a}},\quad
\vartheta^{\hat{c}}=\pi_{\phantom{a}\hat{b}}^{\hat{c}}\,\breve{\vartheta}{}^{\hat{b}},\quad
e\,\vartheta=e_{\hat{a}}\otimes\vartheta^{\hat{a}}=\Omega^{\hat{a}}_{\phantom{a}\hat{b}}\,\breve{e}_{\hat{a}}\otimes
\breve{\vartheta}{}^{\hat{b}}.
\end{array}
\label{R330}
\end{equation}
The matrices
$\pi(\widetilde{\eta}):\phantom{a}=(\pi^{\phantom{a}\hat{a}}_{\hat{b}})(\widetilde{\eta})$
are referred to as the {\em first deformation matrices} and the
matrices $
\gamma_{\hat{c}\hat{d}}(\widetilde{\eta})={}^{*}o_{\hat{a}\hat{b}}\,\pi^{\phantom{a}\hat{a}}_{\hat{c}}
(\widetilde{\eta})\,\pi_{\hat{d}}^{\phantom{a}\hat{b}}(\widetilde{\eta}),
$ \,-\, {\em second deformation matrices}. The matrices $
\pi_{\phantom{a}\hat{c}}^{\hat{a}}(\widetilde{\eta})\,\in\, GL(2,
R)\,\forall\, \widetilde{\eta}, $ in general, give rise to right
cosets of the Lorentz group, i.e. they are the elements of the
quotient group $GL(2, R)/SO(1,1)$, because the Lorentz matrices,
$\Lambda^{r}_{s}$,
 $(r,s=1,0)$ leave the Minkowski metric invariant. A right-multiplication of
$\pi(\widetilde{\eta})$ by a Lorentz matrix gives an other
deformation matrix. So, all the fundamental geometrical structures
on deformed/distorted MS in fact - the metric as much as the
coframes and connections - acquire a {\em deformation/distortion}
induced theoretical interpretation. If we deform the tetrad
according to (\ref{R330}), in general, we may recast metric as
follows:
\begin{equation}
\begin{array}{l}
g={}^{*}o_{\hat{a}\hat{b}}\,\pi_{\phantom{a}\hat{c}}^{\hat{a}}\pi_{\,\,\,\hat{d}}^{\hat{b}}\breve{\vartheta}{}^{\hat{c}}\otimes
\breve{\vartheta}{}^{\hat{d}}=
\gamma_{\hat{c}\hat{d}}\,\breve{\vartheta}{}^{\hat{c}}\otimes
\breve{\vartheta}{}^{\hat{d}}={}^{*}o_{\hat{a}\hat{b}}\,\vartheta^{\hat{a}}\otimes
\vartheta^{\hat{b}}.
\end{array}
\label{R34}
\end{equation}
The deformed metric can be split as \cite{gago1}:
\begin{equation}
\begin{array}{l}
g_{\underline{\mu}\underline{\nu}}(\pi)=\Upsilon^{2}(\pi)\,\breve{g}_{\underline{\mu}\underline{\nu}}+
\gamma_{\underline{\mu}\underline{\nu}}(\pi),
\end{array}
\label{R38}
\end{equation}
where $\Upsilon(\pi)=\pi^{\hat{a}}_{\hat{a}}$, and
\begin{equation}
\begin{array}{l}
\gamma_{\underline{\mu}\underline{\nu}}(\pi)=[\gamma_{\hat{a}\hat{b}}-\Upsilon^{2}(\pi)\,{}^{*}o_{\hat{a}\hat{b}}]\,
\breve{e}{}^{\hat{a}}_{\phantom{a}\underline{\mu}}\,\breve{e}{}^{\hat{b}}_{\phantom{a}\underline{\nu}}.
\end{array}
\label{Rg}
\end{equation}

\subsection{Model building in the 4D background Minkowski space-time}
Here we briefly discuss the RTI in particular case when the
relativistic test particle accelerated in the background flat
$M_{4}$ space under an unbalanced net force other than
gravitational, but we refer to the original paper by ~\cite{gago1}
for more details. To make the remainder of our discussion a bit more
concrete, it proves necessary to provide, further, a constitutive
ansatz of simple, yet tentative, linear {\em distortion
transformations} of the basis
$\underline{e}_{\,\underline{m}}$~(\ref{S3}) at the point of
interest in flat space $\underline{M}_{\,2}$, which can be written
in terms of {\em local rate} $\varrho(\eta,m,f)$ of instantaneously
change of the measure $v^{A}$ of massive $(m)$ test particle under
the unbalanced net force $(f)$~\cite{gago1}:
\begin{equation}
\begin{array}{l}
e_{\tilde{(+)}}(\varrho)=D^{\phantom{a}
\underline{m}}_{\tilde{(+)}}(\varrho)\,
\underline{e}_{\,\underline{m}}=\underline{e}_{\,(+)}-\varrho(\eta,m,f)\, v^{(-)}\,\underline{e}_{\,(-)},\\
e_{\tilde{(-)}}(\varrho)=D^{\phantom{a}
\underline{m}}_{\tilde{(-)}}(\varrho)\,
\underline{e}_{\,\underline{m}}=\underline{e}_{\,(-)}+\varrho(\eta,m,f)\,
v^{(+)}\,\underline{e}_{\,(+)},
\end{array}
\label{R35}
\end{equation}
Clearly, these transformations imply a violation of relation
(\ref{S3})\, ($e_{\underline{\mu}}^{2}(\varrho)\neq 0$) for the null
vectors $\underline{e}_{\,\underline{m}}$. Whereas we simplify
distortion matrices for further use by imposing the constraints
\begin{equation}
\begin{array}{l}
D_{\underline{\mu}}^{\underline{m}}=\breve{D}
{}_{\underline{\mu}}^{\underline{\underline{m}}}\,\,,\quad
\breve{Y}{}^{\underline{\mu}}_{\underline{m}}=\breve{D}
{}^{\underline{\mu}}_{\underline{m}}\,\,,
\end{array}
\label{D6}
\end{equation}
which yields the partial local tetrad deformations
\begin{equation}
\begin{array}{l}
e_{\hat{c}}=\breve{e}_{\hat{c}},\quad
\vartheta^{\hat{c}}=\Omega_{\phantom{a}\hat{b}}^{\hat{c}}\,\breve{\vartheta}{}^{\hat{b}},\quad
e\,\vartheta=e_{\hat{a}}\otimes\vartheta^{\hat{a}}=\Omega^{\hat{a}}_{\phantom{a}\hat{b}}\,\breve{e}_{\hat{a}}
\otimes\breve{\vartheta}{}^{\hat{b}}.
\end{array}
\label{D9}
\end{equation}
The relation (\ref{S5}) now can be rewritten in terms of space-time
variables as
\begin{equation}
\begin{array}{l}
id=e\,\vartheta\equiv d\widetilde{\hat{q}}=\widetilde{e}_{0}\otimes
d\widetilde{t} + \widetilde{e}_{q}\otimes d\widetilde{q},
\label{RE4}
\end{array}
\end{equation}
where $\widetilde{e}_{0}$ and $\widetilde{e}_{q}$ are, respectively,
the temporal and spatial basis vectors:
\begin{equation}
\begin{array}{l}
\widetilde{e}_{0}(\varrho)=
\frac{1}{\sqrt{2}}\left[e_{\tilde{(+)}}(\varrho)+e_{\tilde{(-)}}(\varrho)\right],\quad
\widetilde{e}_{q}(\varrho)=
\frac{1}{\sqrt{2}}\left[e_{\tilde{(+)}}(\varrho)-e_{\tilde{(-)}}(\varrho)\right].
\end{array}
\label{O4}
\end{equation}
Hence, in the framework of the space-time deformation/distortion
theory~\cite{gago1}, we can compute the  general metric
$g$~(\ref{R34}) in $\underline{\widetilde{M}}_{\,2}$ as
\begin{equation}
\begin{array}{l}
 g=g_{\tilde{r}\tilde{s}}\,d\widetilde{q}{}^{\tilde{r}}\otimes
d\widetilde{q}{}^{\tilde{s}},
\end{array}
\label{Tau2}
\end{equation}
provided
\begin{equation}
\begin{array}{l}
 g_{\tilde{0}\tilde{0}}=(1+\frac{\varrho
v_{q}}{\sqrt{2}})^{2}-\frac{\varrho^{2}}{2},\quad
g_{\tilde{1}\tilde{1}}=-(1-\frac{\varrho
v_{q}}{\sqrt{2}})^{2}+\frac{\varrho^{2}}{2},\quad
g_{\tilde{1}\tilde{0}}=g_{\tilde{0}\tilde{1}}=-\sqrt{2}\varrho.
\label{RE494}
\end{array}
\end{equation}
We suppose that a second observer, who makes measurements using a
frame of reference $\widetilde{S}_{(2)}$ which is held stationary in
curved (deformed/distorted) master space
$\widetilde{\mathcal{M}}_{2}$, uses for the test particle the
corresponding space-time coordinates
$\widetilde{q}{}^{\tilde{r}}\left((\widetilde{q}{}^{\tilde{0}},\,
\widetilde{q}{}^{\tilde{1}})\equiv (\widetilde{t},\,\widetilde{q}
)\right)$. The very concept of the local {\em absolute acceleration}
(in Newton's terminology) is introduced by~\cite{gago1}, brought
about via the Fermi-Walker transported frames as
\begin{equation}
\begin{array}{l}
\vec{a}_{abs}\equiv
\vec{e}_{q}\frac{d(\varrho)}{\sqrt{2}ds_{q}}=\vec{e}_{q}\,|\frac{de_{\hat{0}}}{ds}|=\vec{e}_{q}\,|{\bf
a}|.
\end{array}
\label{RL888}
\end{equation}
Here we choose the system $S_{(2)}$ in such a way as the axis
$\vec{e}_{q}$ lies along the net 3-acceleration
($\vec{e}_{q}\,||\,\vec{e}_{a}),\quad
(\vec{e}_{a}=\vec{a}_{net}/|\vec{a}_{net}|$), $\vec{a}_{net}$ is the
local net 3-acceleration of an arbitrary observer with proper linear
3-acceleration $\vec{a}$ and proper 3-angular velocity
$\vec{\omega}$ measured in the rest frame: $
\vec{a}_{net}=\frac{d\vec{u}}{ds}=\vec{a}\,\wedge\,\vec{u}
+\vec{\omega}\times \vec{u},$  where ${\bf u}$ is the 4-velocity. A
magnitude of $\vec{a}_{net}$ can be computed as the simple invariant
of the absolute value $|\frac{d{\bf u}}{ds}|$ as measured in rest
frame:
\begin{equation}
\begin{array}{l}
|{\bf a}|=|\frac{d{\bf u}}{ds}|=
\left(\frac{du^{l}}{ds},\,\frac{du_{l}}{ds}\right)^{1/2}.
\end{array}
\label{G1}
\end{equation}
Also, following \cite{Syn, MTW}, we define an orthonormal frame
$e_{\hat{a}}$, carried by an accelerated observer, who moves with
proper linear 3-acceleration and $\vec{a}(s)$ and proper 3-rotation
$\vec{\omega}(s)$.  Particular frame components are $e_{\hat{a}}$,
where $\hat{a}=\hat{0},\hat{1},$ etc. Let the zeroth leg of the
frame $e_{\hat{0}}$ be 4-velocity ${\bf u}$ of the observer that is
tangent to the worldline at a given event $x^{l}(s)$ and we
parameterize the remaining  spatial triad frame vectors
$e_{\hat{i}}$, orthogonal to $e_{\hat{0}}$, also by $(s)$. The
spatial triad $e_{\hat{i}}$ rotates with proper 3-rotation
$\vec{\omega}(s)$. The 4-velocity vector naturally undergoes
Fermi-Walker transport along the curve C, which guarantees that
$e_{\hat{0}}(s)$ will always be tangent to C determined by $x^{l} =
x^{l}(s)$:
\begin{equation}
\begin{array}{l}
\frac{de_{\hat{a}}}{ds}=-\Phi\, e_{\hat{a}}
\end{array}
\label{G2}
\end{equation}
where the antisymmetric rotation tensor $\Phi$ splits into a
Fermi-Walker transport part  $\Phi_{FW}$ and a spatial rotation part
$\Phi_{SR}$:
\begin{equation}
\begin{array}{l}
\Phi^{lk}_{FW}=a^{l}u^{k}-a^{k}u^{l},\quad
\Phi^{lk}_{SR}=u_{m}\omega_{n}\varepsilon^{mnlk}.
\end{array}
\label{G3}
\end{equation}
The 4-vector of rotation $\omega^{l}$ is orthogonal to 4-velocity
$u^{l}$, therefore, in the rest frame it becomes
$\omega^{l}(0,\,\vec{\omega})$, and $\varepsilon^{mnlk}$ is the
Levi-Civita tensor with $\varepsilon^{0123}=-1$. So, the resulting
metric (\ref{Tau2}) is reduced to
\begin{equation}
\begin{array}{l}
 d\widetilde{s}{}^{2}_{q}=
\Omega^{2}(\overline{\varrho})\,ds^{2}_{q},\quad
\Omega(\overline{\varrho})=1+\overline{\varrho}{}^{2}$,
$\,\overline{\varrho}{}^{2}=v^{2}\varrho^{2},\quad
v^{2}=v^{(+)}v^{(-)},\quad \varrho=\sqrt{2}\,\int^{s_{q}}_{0}|{\bf
a}|ds_{q}'.
\end{array}
\label{W7}
\end{equation}
Combining ~(\ref{T1}) and ~(\ref{RL888}), we obtain
\begin{equation}
\begin{array}{l}
\varrho=\frac{i}{\gamma_{q}^{2}}\left|\left(\underline{\theta}\,\sigma^{3}\frac{d\bar{\xi}}{ds_{q}}-
\frac{d\xi}{ds_{q}}\sigma^{3}\underline{\bar{\theta}}\right)\right|,
\end{array}
\label{LL1}
\end{equation}
where $ \gamma_{q}=(1-v_{q}^{2})^{-1/2}$.  The resulting {\em
inertial force} $\vec{f}_{(in)}$ is computed by~\cite{gago1} as
\begin{equation}
\begin{array}{l}
\vec{f}_{(in)}=-m\,
\Gamma^{1}_{\tilde{r}\tilde{s}}(\varrho)\frac{d\widetilde{q}{}^{\tilde{r}}}{d\widetilde{s}_{q}}
\frac{d\widetilde{q}^{\tilde{s}}}{d\widetilde{s}_{q}}=
-\frac{m\vec{a}_{abs}}{\Omega^{2}(\overline{\varrho})\,\gamma_{q}},
\end{array}
\label{G33}
\end{equation}
Whereupon, in case of absence of rotation,  the relativistic
inertial force reads
\begin{equation}
\begin{array}{l}
\vec{f}_{(in)}=-\frac{1}{\Omega^{2}(\overline{\varrho})\,\gamma_{q}\gamma}[\vec{F}+(\gamma-1)
\frac{\vec{v}(\vec{v}\cdot\vec{F})}{|\vec{v}|^{2}}].
\end{array}
\label{RLL999}
\end{equation}
Note that the inertial force arises due to {\it nonlinear} process
of deformation of MS,  resulting after all to {\it linear} relation
~(\ref{RLL999}). So, this also ultimately requires that MS should be
two dimensional, because  in this case we may reconcile the alluded
{\it nonlinear} and {\it linear} processes by choosing the system
$S_{(2)}$ in only allowed way mentioned above. At low velocities
$v_{q}\simeq |\vec{v}|\simeq 0$ and tiny accelerations we usually
experience, one has $\Omega(\overline{\varrho})\simeq 1$, therefore
the ~(\ref{RLL999}) reduces to the conventional non-relativistic law
of inertia
\begin{equation}
\begin{array}{l}
\vec{f}_{(in)}=-m\vec{a}_{abs}=-\vec{F}.
\end{array}
\label{RLG7}
\end{equation}
At high velocities $v_{q}\simeq|\vec{v}|\simeq 1$
\,($\Omega(\overline{\varrho})\simeq 1$), if
$(\vec{v}\cdot\vec{F})\neq 0,$  the inertial force (\ref{RLL999})
becomes
\begin{equation}
\begin{array}{l}
\vec{f}_{(in)}\simeq-\frac{1}{\gamma}\vec{e}_{v}(\vec{e}_{v}\cdot\vec{F}),
\end{array}
\label{RKK}
\end{equation}
and it vanishes in the limit of the photon $(|\vec{v}|= 1, \,m=0).$
Thus, it takes force to disturb an inertia state, i.e. to make the
{\em absolute acceleration} ($\vec{a}_{abs}\neq 0$). The {\em
absolute acceleration} is due to the  real deformation/distortion of
the space $\underline{M}_{\,2}$. The {\em relative}
($d(\tau_{2}\varrho)/ds_{q}=0$) acceleration (in Newton's
terminology) (both magnitude and direction), to the contrary, has
nothing to do with the deformation/distortion of the space
$\underline{M}_{\,2}$ and, thus, it cannot produce an inertia
effects.

\subsection{Beyond the hypothesis of locality}
In SR an assumption is required to relate the ideal inertial
observers to actual observers that are all noninertial, i.e.,
accelerated.  Therefore, it is a long-established practice in
physics to use the hypothesis of locality \cite{MTW}-\cite{MB8}),
for extension of the Lorentz invariance to accelerated observers in
Minkowski space-time. The standard geometrical structures, referred
to a noninertial coordinate frame of accelerating and rotating
observer in Minkowski space-time, were computed on the base of the
assumption that an accelerated observer is pointwise inertial, which
in effect replaces an accelerated observer at each instant with a
momentarily comoving inertial observer along its wordline. This
assumption is known to be an approximation limited to motions with
sufficiently low accelerations, which works out because all relevant
length scales in feasible experiments are very small in relation to
the huge acceleration lengths of the tiny accelerations we usually
experience, therefore, the curvature of the wordline could be
ignored and that the differences between observations by accelerated
and comoving inertial observers will also be very small. However, it
seems quite clear that such an approach is a work in progress, which
reminds us of a puzzling underlying reality of inertia, and that it
will have to be extended to describe physics for arbitrary
accelerated observers. Ever since this question has become a major
preoccupation of physicists, see e.g. \cite{HL}-\cite{Marz} and
references therein. The hypothesis of locality  represents strict
restrictions, because in other words, it approximately replaces a
noninertial frame of reference $\widetilde{S}_{(2)}$, which is held
stationary in the deformed/distorted space $\widetilde{M}_{2}\equiv
\underline{V}_{\,2}^{(\varrho)}\, (\varrho\neq 0)$, with a
continuous infinity set of the inertial frames
$\{S_{(2)},\,S'_{(2)},\,S_{(2)}'',...\}$ given in the flat
$\underline{M}_{2}\, (\varrho=0)$. In this situation the use of the
hypothesis of locality is physically unjustifiable. Therefore, it is
worthwhile to go beyond the hypothesis of locality with special
emphasis on distortion of
$\underline{M}_{\,2}$\,\,($\underline{M}_{2}\longrightarrow
\underline{V}_{\,2}^{(\varrho)}$), which we might expect will
essentially improve the standard results. The notation will be
slightly different from the previous subsection. We now denote the
orthonormal frame $e_{\hat{a}}$ (\ref{G2}), carried by an
accelerated observer, with the over '{\em breve}' such that
\begin{equation}
\begin{array}{l}
\breve{e}_{\hat{a}}=\overline{e}{}_{\phantom{a}\hat{a}}^{\,
\mu}\,\overline{e}_{\mu}=
\breve{e}{}_{\phantom{a}\hat{a}}^{\,\mu}\,\breve{e}_{\mu},\quad
\breve{\vartheta}{}^{\hat{b}}=\overline{e}{}^{\phantom{a}\hat{b}}_{
\mu}\,\overline{\vartheta}{}^{\mu}=
\breve{e}{}^{\phantom{a}\hat{b}}_{\,\mu}\,\breve{\vartheta}^{\mu},
\end{array}
\label{W01}
\end{equation}
with $\overline{e}_{\mu}=\partial_{\mu}=\partial/\partial
x^{\mu},\quad
\breve{e}_{\mu}=\breve{\partial}_{\mu}=\partial/\partial
\breve{x}{}^{\mu},\quad \overline{\vartheta}{}^{\mu}=dx^{\mu},\quad
\breve{\vartheta}^{\mu}=d\breve{x}. $ Here, following \cite{MTW,
MB1}, we introduced a geodesic coordinate system
$\breve{x}{}^{\mu}$\,-\, "coordinates relative to the accelerated
observer" (laboratory coordinates), in the neighborhood of the
accelerated path. The coframe members
$\{\breve{\vartheta}{}^{\,\hat{b}}\}$ are the objects of dual
counterpart:\,
$\breve{e}_{\hat{a}}\,\rfloor\,\breve{\vartheta}{}^{\hat{b}}=\delta^{b}_{a}$.
We choose the zeroth leg of the frame, $\breve{e}_{\hat{0}}$, as
before, to be the unit vector ${\bf u}$ that is tangent to the
worldline at a given event $x^{\mu}(s)$, where $(s)$ is a proper
time measured along the accelerated path by the standard (static
inertial) observers in the underlying global inertial frame. The
condition of orthonormality for the frame field
$\overline{e}{}^{\,\mu}_{\phantom{a}\hat{a}}$ reads
$\eta_{\mu\nu}\,\overline{e}{}^{\,\mu}_{\phantom{a}\hat{a}}\,\overline{e}{}^{\,\nu}_{\phantom{a}\hat{b}}=
o_{\hat{a}\hat{b}}=diag(+---)$. The antisymmetric acceleration
tensor $\Phi_{ab}$ \cite{MB1}-\cite{MF} is given by
\begin{equation}
\begin{array}{l}
\Phi_{a}^{\phantom{a}b}:\,=\overline{e}{}^{\phantom{a}\hat{b}}_{
\mu}\,\frac{d\overline{e}{}_{\phantom{a}\hat{a}}^{\,
\mu}}{ds}=\overline{e}{}^{\phantom{a}\hat{b}}_{
\mu}\,u^{\lambda}\,\breve{\nabla}_{\lambda}\,\overline{e}{}_{\phantom{a}\hat{a}}^{\,
\mu}=u\,\rfloor\,\breve{\Gamma}_{a}^{\phantom{a}b},
\end{array}
\label{WW1}
\end{equation}
provided
$\breve{\Gamma}_{a}^{\phantom{a}b}=\breve{\Gamma}{}_{a\mu}^{\phantom{a}b}\,d\breve{x}{}^{\mu}$,
where $\breve{\Gamma}{}_{a\mu}^{\phantom{a}b}$ is the metric
compatible, torsion-free Levi-Civita connection. According to
(\ref{G2}) and (\ref{G3}), and in analogy with the Faraday tensor,
one can identify $\Phi_{ab}\longrightarrow (-{\bf a},\,{\bf
\omega})$, with ${\bf a}(s)$ as the translational acceleration
$\Phi_{0i}=-a_{i}$, and ${\bf \omega}(s)$ as the frequency of
rotation of the local spatial frame with respect to a nonrotating
(Fermi- Walker transported) frame
$\Phi_{ij}=-\varepsilon_{ijk}\,\omega^{k}$. The invariants
constructed out of $\Phi_{ab}$ establish the acceleration scales and
lengths. The hypothesis of locality holds for huge proper
acceleration lengths $|I|^{-1/2}\gg 1$ and $|I^{*}|^{-1/2}\gg 1$,
where the scalar invariants are given by
\,$I=(1/2)\,\Phi_{ab}\,\Phi^{ab}=-\vec{a}^{2}+\vec{\omega}^{2}$ and
$I^{*}=(1/4)\,\Phi_{ab}^{*}\,\Phi^{ab}=-\vec{a}\cdot \vec{\omega}$
($\Phi_{ab}^{*}=\varepsilon_{abcd}\,\Phi^{cd}$) \cite{MB1,
MB2,MB5,MB6,MB7,MB8}. Suppose the displacement vector $z^{\mu}(s)$
represents the position of the accelerated observer. According to
the hypothesis of locality, at any time $(s)$ along the accelerated
worldline the hypersurface orthogonal to the worldline is Euclidean
space and we usually describe some event on this hypersurface
("local coordinate system") at $x^{\mu}$ to be at
$\breve{x}{}^{\mu}$, where $x^{\mu}$ and $\breve{x}^{\mu}$ are
connected via $\breve{x}{}^{\,0}= s$ and
\begin{equation}
\begin{array}{l}
x^{\mu}=z^{\mu}(s)+\breve{x}{}^{\,i}\,\overline{e}{}^{\,
\mu}_{\phantom{a}\hat{i}}(s).
\end{array}
\label{W1}
\end{equation}
Let $\breve{q}{}^{\,r}(\breve{q}{}^{\,0},\, \breve{q}{}^{\,1})$ be
"coordinates relative to the accelerated observer" in the
neighborhood of the accelerated path in $\underline{M}_{\,2}$, with
space-time components implying
\begin{equation}
\begin{array}{l}
d\breve{q}{}^{\,0}=d\breve{x}{}^{\,0},\quad
d\breve{q}{}^{\,1}=|d\vec{\breve{x}}|,\quad
\vec{\breve{e}}=\frac{d\vec{\breve{x}}}{d\breve{q}{}^{\,1}}=\frac{d\vec{\breve{x}}}{|d\vec{\breve{x}}|},\quad
\vec{\breve{e}}\cdot\vec{\breve{e}}=1.
\end{array}
\label{W02}
\end{equation}
As long as a locality assumption holds, we may describe, with equal
justice, the event at $x^{\mu}$ (\ref{W1}) to be at point
$\breve{q}{}^{\,r}$, such that $x^{\mu}$ and $\breve{q}{}^{\,r}$, in
full generality, are connected via $\breve{q}{}^{\,0}=s$ and
\begin{equation}
\begin{array}{l}
x^{\mu}=z^{\mu}_{q}(s)+\breve{q}{}^{\,1}\,\overline{\beta}{}^{\mu}_{\phantom{a}\hat{1}}(s),
\end{array}
\label{W2}
\end{equation}
where the displacement vector from the origin reads
$dz^{\mu}_{q}(s)=\overline{\beta}{}^{\,\mu}_{\phantom{a}\hat{0}}\,d\breve{q}{}^{\,0}$,
and the components $\overline{\beta}{}^{\,\mu}_{\phantom{a}\hat{r}}$
can be written in terms of
$\overline{e}{}_{\phantom{a}\hat{a}}^{\,\mu}$. Actually, from
(\ref{W1}) and (\ref{W2}) we may obtain
\begin{equation}
\begin{array}{l}

dx^{\mu}=dz^{\mu}_{q}(s)+d\breve{q}{}^{\,1}\,\overline{\beta}{}^{\,\mu}_{\phantom{a}\hat{1}}(s)+
\breve{q}{}^{\,1}\,d\overline{\beta}^{\,\mu}_{\phantom{a}\hat{1}}(s)=
\left[\overline{\beta}{}^{\,\mu}_{\phantom{a}\hat{0}}(1+
\breve{q}{}^{\,1}\check{\varphi}_{0})+\overline{\beta}{}^{\,\mu}_{\phantom{a}\hat{1}}\,\breve{q}{}^{\,1}\check{\varphi}_{1}\right]d\breve{q}{}^{\,0}+
\overline{\beta}{}^{\,\mu}_{\phantom{a}\hat{1}}\,d\breve{q}{}^{\,1}\\
\equiv
dz^{\mu}(s)+d\breve{x}{}^{\,i}\,\overline{e}{}^{\,\mu}_{\phantom{a}\hat{i}}(s)+
\breve{x}{}^{\,i}\,d\overline{e}{}^{\,\mu}_{\phantom{a}\hat{i}}(s)=\left[\overline{e}{}^{\,\mu}_{\phantom{a}\hat{0}}(1+
\breve{x}{}^{\,i}\Phi^{0}_{i})+\overline{e}{}^{\,\mu}_{\phantom{a}\hat{j}}\,\breve{x}{}^{\,i}\Phi^{j}_{i})\right]\,
d\breve{x}{}^{\,0}+
\overline{e}{}^{\,\mu}_{\phantom{a}\hat{i}}\,d\breve{x}{}^{\,i},
\end{array}
\label{W3}
\end{equation}
where $d\overline{\beta}{}^{\,\mu}_{\phantom{a}\hat{1}}(s)$ is
written in the basis
$\overline{\beta}{}^{\,\mu}_{\phantom{a}\hat{a}}$ as
$d\overline{\beta}{}^{\,\mu}_{\phantom{a}\hat{1}}=(\check{\varphi}_{0}\overline{\beta}{}^{\,\mu}_{\phantom{a}\hat{0}}+
\check{\varphi}_{1}\overline{\beta}{}^{\,\mu}_{\phantom{a}\hat{1}})d\breve{q}{}^{\,0}$.
The equation (\ref{W3}) holds by identifying
\begin{equation}
\begin{array}{l}
 \overline{\beta}{}^{\,\mu}_{\phantom{a}\hat{0}}\left(1+
\breve{q}{}^{\,1}\breve{\varphi}_{0}\right)\equiv
\overline{e}{}^{\,\mu}_{\phantom{a}\hat{0}}\left(1+
\breve{x}{}^{\,i}\Phi^{0}_{i}\right),\quad
\overline{\beta}{}^{\,\mu}_{\phantom{a}\hat{1}}\,\breve{q}{}^{\,1}\breve{\varphi}_{1}\equiv
\overline{e}{}^{\,\mu}_{\phantom{a}\hat{j}}\,\breve{x}{}^{\,i}\Phi^{j}_{i},\quad
\overline{\beta}{}^{\,\mu}_{\phantom{a}\hat{1}}\,d\breve{q}{}^{\,1}\equiv
\overline{e}{}^{\,\mu}_{\phantom{a}\hat{i}}\,d\breve{x}{}^{\,i}.
\end{array}
\label{W4}
\end{equation}
Choosing $\overline{\beta}{}^{\,\mu}_{\phantom{a}\hat{0}}\equiv
\overline{e}{}^{\,\mu}_{\phantom{a}\hat{0}}$, we have then
\begin{equation}
\begin{array}{l}
\breve{q}{}^{\,1}\breve{\varphi}_{0}=
\breve{x}{}^{\,i}\,\Phi^{0}_{i},\quad
\overline{\beta}{}^{\,\mu}_{\phantom{a}\hat{1}}=\overline{e}{}^{\,\mu}_{\phantom{a}\hat{i}}\,\breve{e}{}^{\,i},\quad
\breve{q}{}^{\,1}\breve{\varphi}_{1}
=\breve{x}{}^{\,i}\,\Phi^{j}_{i}\,\breve{e}{}^{\,-1}_{j},
\end{array}
\label{W5}
\end{equation}
with $\breve{e}{}^{j}\,\breve{e}{}^{\,-1}_{i}=\delta^{j}_{i}$.
Consequently, (\ref{W3}) yields the standard metric of
semi-Riemannian 4D background space $V^{(0)}_{4}$ in noninertial
system of the accelerating and rotating observer, computed on the
base of hypothesis of locality:
\begin{equation}
\begin{array}{l}
  \breve{g}=\eta_{\mu\nu}\,dx^{\mu}\otimes dx^{\nu}=\left[(1+
\vec{a}\cdot
\vec{\breve{x}})^{2}+(\vec{\omega}\cdot\vec{\breve{x}})^{2}-(\vec{\omega}\cdot\vec{\omega})(\vec{\breve{x}}\cdot
\vec{\breve{x}})\right]\,d\breve{x}{}^{0}\otimes d\breve{x}{}^{0}-\\
2\,(\vec{\omega}\wedge\vec{\breve{x}}) \cdot d\vec{\breve{x}}\otimes
d\breve{x}{}^{0}- d\vec{\breve{x}}\otimes d\vec{\breve{x}},
\end{array}
\label{W19}
\end{equation}
This metric was derived by \cite{Hehl7} and \cite{HL}, in agreement
with \cite{Ni77} -\cite{Ni78D} (see also
\cite{MB1,MB2,MB5,MB6,MB7,MB8}). We see that the hypothesis of
locality leads to the 2D semi-Riemannian space,
$\underline{V}^{(0)}_{\,2}$, with the incomplete metric
$\breve{g}\quad(\varrho=0)$:
\begin{equation}
\begin{array}{l}
 \breve{g}=\left[(1+
\breve{q}{}^{\,1}\breve{\varphi}_{0})^{2}-(\breve{q}{}^{\,1}\breve{\varphi}_{1})^{2}\right]\,d\breve{q}{}^{\,0}\otimes
d\breve{q}{}^{\,0}-  2\,(\breve{q}{}^{\,1}\breve{\varphi}_{1})\,
d\breve{q}{}^{\,1}\otimes d\breve{q}{}^{\,0}-
d\breve{q}{}^{\,1}\otimes d\breve{q}{}^{\,1},
\end{array}
\label{W05}
\end{equation}
Therefore, our strategy now is to deform the metric (\ref{W05}) by
carrying out an additional deformation of semi-Riemannian 4D
background space $V^{(0)}_{4}\longrightarrow \widetilde{M}_{4}\equiv
V^{(\varrho)}_{4}$, in order it becomes on the same footing with the
complete metric $g\quad(\varrho\neq 0)$ (\ref{Tau2}) of the
distorted space $\underline{\widetilde{M}}_{\,2}\,\equiv\,
\underline{V}^{(\varrho)}_{\,2}$.  Let the Latin letters
$\hat{r},\hat{s},...=0,1$ be the anholonomic indices referred to the
anholonomic frame
$e_{\hat{r}}=e_{\phantom{a}\hat{r}}^{s}\,\partial_{\tilde{s}}$,
defined on the $\underline{V}^{(\varrho)}_{\,2}$, with
$\partial_{\tilde{s}}=\partial/\partial\,\widetilde{q}{}^{\tilde{s}}$
as the vectors tangent to the coordinate lines. So, a smooth
differential 2D-manifold $\underline{V}^{(\varrho)}_{\,2}$ has at
each point $\widetilde{q}^{s}$ a tangent space
$\widetilde{T}_{\widetilde{q}}\underline{V}^{(\varrho)}_{\,2}$,
spanned by the frame, $\{e_{\hat{r}}\}$, and the coframe members
$\vartheta^{\hat{r}}=e^{\phantom{a}\hat{r}}_{s}\,d\widetilde{q}{}^{\tilde{s}}$,
which constitute a basis of the covector space
$\widetilde{T}{}^\star_{\widetilde{q}}\underline{V}^{(\varrho)}_{\,2}$.
All this nomenclature can be given for $\underline{V}^{(0)}_{\,2}$
too. Then, we may compute corresponding vierbein fields
$\breve{e}{}^{\phantom{a}\hat{s}}_{r}$ and
$e^{\phantom{a}\hat{s}}_{r}$  from the equations
\begin{equation}
\begin{array}{l}
\breve{g}_{rs}=\breve{e}{}^{\phantom{a}\hat{r'}}_{r}\,\breve{e}{}^{\phantom{a}\hat{s'}}_{s}\,o_{\hat{r'}\hat{s'}},\quad
g_{\tilde{r}\tilde{s}}(\varrho)=e^{\phantom{a}\hat{r'}}_{r}(\varrho)\,e^{\phantom{a}\hat{s'}}_{s}(\varrho)\,o_{\hat{r'}\hat{s'}},
\end{array}
\label{W09}
\end{equation}
with  $\breve{g}_{rs}$ (\ref{W05}) and
$g_{\tilde{r}\tilde{s}}(\varrho)$ (\ref{RE494}). Hence
\begin{equation}
\begin{array}{l}
 \breve{e}{}^{\phantom{a}\hat{0}}_{0}=1+ \vec{a}\cdot
\vec{\breve{x}},\quad
\breve{e}{}^{\phantom{a}\hat{1}}_{0}=\vec{\omega}\,\wedge\,\vec{\breve{x}},\quad
\breve{e}{}^{\phantom{a}\hat{0}}_{1}=0,\quad
\breve{e}{}^{\phantom{a}\hat{1}}_{1}=1,\\
 e^{\phantom{a}\hat{0}}_{0}(\varrho)=1+\frac{\varrho
v_{q}}{\sqrt{2}},\quad
e^{\phantom{a}\hat{1}}_{0}(\varrho)=\frac{\varrho}{\sqrt{2}},\quad
e^{\phantom{a}\hat{0}}_{1}(\varrho)=-\frac{\varrho}{\sqrt{2}},\quad
e^{\phantom{a}\hat{1}}_{1}(\varrho)=1-\frac{\varrho
v_{q}}{\sqrt{2}}.
\end{array}
\label{W10}
\end{equation}
Since a distortion $\underline{M}_{\,2}\longrightarrow
\underline{\widetilde{M}}_{\,2}$ may affect only the
$\underline{M}_{\,2}$-part of the components
$\overline{\beta}{}^{\,\mu}_{\phantom{a}\hat{r}}$, without relation
to the 4D background space-time part, therefore, a deformation
$V^{(0)}_{4}\longrightarrow V^{(\varrho)}_{4}$  is equivalent to a
straightforward generalization
$\overline{\beta}{}^{\,\mu}_{\phantom{a}\hat{r}}\longrightarrow
\beta^{\mu}_{\phantom{a}\hat{r}}(\varrho)$, where
\begin{equation}
\begin{array}{l}
\beta^{\mu}_{\phantom{a}\hat{r}}(\varrho)=
E^{\phantom{a}\hat{s}}_{\hat{r}}(\varrho)\,\overline{\beta}{}^{\,\mu}_{\phantom{a}\hat{s}},\quad
E^{\phantom{a}\hat{s}}_{\hat{r}}(\varrho):\phantom{a}=
e^{r'}_{\phantom{a}\hat{r}}(\varrho)\,\breve{e}{}^{\phantom{a}\hat{s}}_{r'}.
\end{array}
\label{W6}
\end{equation}
Consequently, the (\ref{W6}) gives a generalization of (\ref{W1}) as
\begin{equation}
\begin{array}{l}
x^{\mu}\,\longrightarrow\,x^{\mu}_{(\varrho)}=z^{\mu}_{(\varrho)}(s)+
\breve{x}{}^{\,i}\,e^{\mu}_{\phantom{a}\hat{i}}(s),
\end{array}
\label{W8}
\end{equation}
provided, as before, $\breve{x}{}^{\mu}$ denotes the coordinates
relative to the accelerated observer in 4D background space
$V_{4}^{(\varrho)}$, and according to (\ref{W4}), we have
\begin{equation}
\begin{array}{l}
e^{\mu}_{\phantom{a}\hat{0}}(\varrho)=\beta^{\mu}_{\phantom{a}\hat{0}}(\varrho),\quad
e^{\mu}_{\phantom{a}\hat{i}}(\varrho)=\beta^{\mu}_{\phantom{a}\hat{1}}(\varrho)\,\breve{e}{}^{-1}_{i}.
\end{array}
\label{W9}
\end{equation}
A displacement vector from the origin is then
$dz^{\mu}_{\varrho}(s)=e^{\mu}_{\phantom{a}\hat{0}}(\varrho)\,d\breve{x}{}^{0}$,
Combining (\ref{W6}) and (\ref{W9}), and inverting
$e^{\phantom{a}\hat{s}}_{r}(\varrho)$ (\ref{W10}), we obtain  $
e^{\mu}_{\phantom{a}\hat{a}}(\varrho)=\pi^{\phantom{a}\hat{b}}_{\hat{a}}(\varrho)\,\overline{e}{}^{\,\mu}_{\phantom{a}\hat{b}}$,
where
\begin{equation}
\begin{array}{l}
  \pi^{\hat{0}}_{\hat{0}}(\varrho)\equiv
(1+\frac{\varrho^{2}}{2\gamma^{2}_{q}})^{-1}(1-\frac{\varrho
v_{q}}{\sqrt{2}})\,(1+ \vec{a}\cdot \vec{\breve{x}}),\quad
\pi^{\hat{i}}_{\hat{0}}(\varrho)\equiv
-(1+\frac{\varrho^{2}}{2\gamma^{2}_{q}})^{-1}\frac{\varrho}{\sqrt{2}}\,\breve{e}^{i}\,(1+
\vec{a}\cdot
\vec{\breve{x}}),\\

\pi^{\hat{0}}_{\hat{i}}(\varrho)\equiv(1+\frac{\varrho^{2}}{2\gamma^{2}_{q}})^{-1}\left[(\vec{\omega}\wedge\vec{\breve{x}})(1-\frac{\varrho
v_{q}}{\sqrt{2}})-\frac{\varrho}{\sqrt{2}}\right]\,\breve{e}^{-1}_{i},\quad
\pi^{\hat{j}}_{\hat{i}}(\varrho)=\delta^{j}_{i}\,\pi(\varrho),\\

\pi(\varrho)\equiv(1+\frac{\varrho^{2}}{2\gamma^{2}_{q}})^{-1}\,\left[(\vec{\omega}\wedge\vec{\breve{x}})\,\frac{\varrho}{\sqrt{2}}+
1+\frac{\varrho v_{q}}{\sqrt{2}}\right] .
\end{array}
\label{W12}
\end{equation}
Thus,
\begin{equation}
\begin{array}{l}
dx^{\mu}_{\varrho}=dz^{\mu}_{\varrho}(s)+d\breve{x}{}^{\,i}\,e^{\,\mu}_{\phantom{a}\hat{i}}+
\breve{x}{}^{\,i}\,de^{\,\mu}_{\phantom{a}\hat{i}}(s)=(\tau^{\hat{b}}\,d\breve{x}{}^{0}+
\pi^{\hat{b}}_{\hat{i}}\,d\breve{x}{}^{\,i})\,\overline{e}{}^{\,\mu}_{\phantom{a}\hat{b}}\,,
\end{array}
\label{W13}
\end{equation}
where
\begin{equation}
\begin{array}{l}
\tau^{\hat{b}}\equiv
\pi^{\hat{b}}_{\hat{0}}+\breve{x}{}^{\,i}\,\left(\pi^{\hat{a}}_{\hat{i}}\Phi^{b}_{a}+\frac{d\pi^{\hat{b}}_{\hat{i}}}{ds}\right).
\end{array}
\label{W14}
\end{equation}
Hence, in general, the metric in noninertial frame of arbitrary
accelerating and rotating observer in Minkowski space-time  is
\begin{equation}
\begin{array}{l}
g(\varrho)=\eta_{\mu\nu}\,dx^{\mu}_{\varrho}\otimes
dx^{\nu}_{\varrho}= W_{\mu\nu}(\varrho)\,d\breve{x}{}^{\mu}\otimes
d\breve{x}^{\nu},
\end{array}
\label{W15}
\end{equation}
which can be conveniently decomposed according to
\begin{equation}
\begin{array}{l}
W_{00}(\varrho)= \pi^{2}\left[(1+ \vec{a}\cdot
\vec{\breve{x}})^{2}+(\vec{\omega}\cdot\vec{\breve{x}})^{2}-(\vec{\omega}\cdot\vec{\omega})(\vec{\breve{x}}\cdot
\vec{\breve{x}})\right]+\gamma_{00}(\varrho),\\
 W_{0i}(\varrho)=-\pi^{2}\,(\vec{\omega}\wedge\vec{\breve{x}})^{i}+\gamma_{0i}(\varrho),\quad
W_{ij}(\varrho)=-\pi^{2}\,\delta_{ij}+\gamma_{ij}(\varrho),
\end{array}
\label{W16}
\end{equation}
and that
\begin{equation}
\begin{array}{l}
 \gamma_{00}(\varrho)=\pi\,\left[(1+ \vec{a}\cdot
\vec{\breve{x}})\zeta^{0}-
(\vec{\omega}\wedge\vec{\breve{x}})\cdot\vec{\zeta}\right]+(\zeta^{0})^{2}-(\vec{\zeta})^{2},\quad
 \gamma_{0i}(\varrho)=-\pi\,\zeta^{i}+\tau^{\hat{0}}\,\pi^{\hat{0}}_{\hat{i}},\\

\gamma_{ij}(\varrho)=\pi^{\hat{0}}_{\hat{i}}\,\pi^{\hat{0}}_{\hat{j}},\quad
\zeta^{0}= \pi\,\left(\tau^{\hat{0}}-1- \vec{a}\cdot
\vec{\breve{x}}\right),\quad
\vec{\zeta}=\pi\,\left(\vec{\tau}-\vec{\omega}\wedge\vec{\breve{x}}\right).
\end{array}
\label{W17}
\end{equation}
As we expected, according to (\ref{W15})- (\ref{W17}), the matric
$g(\varrho)$ is decomposed in the form of (\ref{R38}):
\begin{equation}
\begin{array}{l}
g(\varrho)=\pi^{2}(\varrho)\,\breve{g} +\gamma(\varrho),
\end{array}
\label{W015}
\end{equation}
where
$\gamma(\varrho)=\gamma_{\mu\nu}(\varrho)\,d\breve{x}{}^{\mu}\otimes
d\breve{x}^{\nu}$ and
$\Upsilon(\varrho)=\pi^{\hat{a}}_{\hat{a}}(\varrho)= \pi(\varrho)$.
In general, the geodesic coordinates are admissible as long as
\begin{equation}
\begin{array}{l}
\left(1+ \vec{a}\cdot
\vec{\breve{x}}+\frac{\zeta^{0}}{\pi}\right)^{2}\,>\,\left(\vec{\omega}\wedge\vec{\breve{x}}+\frac{\vec{\zeta}}{\pi}\right)^{2}.
\end{array}
\label{W22}
\end{equation}
The equations (\ref{W19}) and (\ref{W15}) say that the vierbein
fields, with entries
$\eta_{\mu\nu}\,\overline{e}{}^{\,\mu}_{\phantom{a}\hat{a}}\,\overline{e}^{\,\nu}_{\phantom{a}\hat{b}}=o_{\hat{a}\hat{b}}$
and
$\eta_{\mu\nu}\,e^{\,\mu}_{\phantom{a}\hat{a}}\,e^{\,\nu}_{\phantom{a}\hat{b}}=\gamma_{\hat{a}\hat{b}}$
lead to the relations
\begin{equation}
\begin{array}{l}

\breve{g}=o_{\hat{a}\hat{b}}\,\breve{\vartheta}{}^{\hat{a}}\otimes\breve{\vartheta}{}^{\hat{b}},\quad
g
=o_{\hat{a}\hat{b}}\,\vartheta^{\hat{a}}\otimes\vartheta^{\hat{b}}=
\gamma_{\hat{a}\hat{b}}\,\breve{\vartheta}{}^{\hat{a}}\otimes\breve{\vartheta}{}^{\hat{b}}=(\Omega^{\phantom{a}
\hat{c}}_{\hat{a}}\,\Omega^{\phantom{a}
\hat{d}}_{\hat{b}}\,o_{\hat{c}\hat{d}})\,\bar{\vartheta}{}^{\hat{a}}\otimes\bar{\vartheta}{}^{\hat{b}},
\end{array}
\label{W012}
\end{equation}
and that (\ref{W3}) and (\ref{W13}) readily give the coframe fields:
\begin{equation}
\begin{array}{l}
\breve{\vartheta}{}^{\hat{b}}=\overline{e}{}_{\,\mu}^{\phantom{a}\hat{b}}\,dx^{\mu}=
\breve{e}{}^{\hat{b}}_{\phantom{a}\mu}\,d\breve{x}{}^{\,\mu},\quad
\breve{e}{}^{\hat{b}}_{\phantom{a}0}=N^{b}_{0},\quad
\breve{e}{}^{\hat{b}}_{\phantom{a}i}=N^{b}_{i},\\
\vartheta^{\hat{b}}=\overline{e}{}_{\,\mu}^{\phantom{a}\hat{b}}\,dx^{\mu}_{\varrho}=
e^{\hat{b}}_{\phantom{a}\mu}\,d\breve{x}{}^{\,\mu}=\pi^{\hat{b}}_{\phantom{a}\hat{a}}\,\breve{\vartheta}{}^{\hat{a}},\quad
e^{\hat{b}}_{\phantom{a}0}=\tau^{\hat{b}},\quad
e^{\hat{b}}_{\phantom{a}i}=\pi^{\hat{b}}_{\phantom{a}\hat{i}}.
\end{array}
\label{W24}
\end{equation}
where $N^{0}_{0}=N\equiv \left(1+ \vec{a}\cdot
\vec{\breve{x}}\right),\quad N^{0}_{i}=0,\quad N^{i}_{0}=N^{i}\equiv
\left(\vec{\omega}\cdot\vec{\breve{x}}\right)^{i},\quad
N^{j}_{i}=\delta^{j}_{i}$. In the standard $(3+1)$-decomposition of
space-time, $N$ and $N^{i}$ are known as {\it lapse function} and
{\it shift vector}, respectively \cite{Gron}. Hence, we may easily
recover the frame field
$e_{\hat{a}}=e_{\phantom{a}\hat{a}}^{\mu}\,\breve{e}_{\mu}=\pi^{\phantom{a}\hat{b}}_{\hat{a}}\,\breve{e}_{\hat{b}}$\,\,
by inverting (\ref{W24}):
\begin{equation}
\begin{array}{l}

e_{\hat{0}}(\varrho)=\frac{\pi(\varrho)}{\pi(\varrho)\,\tau^{\hat{0}}(\varrho)-
\pi^{\hat{0}}_{\hat{k}}(\varrho)\,\tau^{\hat{k}}(\varrho)}\,\breve{e}_{0}\,-\,
\frac{\tau^{\hat{i}}(\varrho)}
{\pi(\varrho)\,\tau^{\hat{0}}(\varrho)-\pi^{\hat{0}}_{\hat{k}}(\varrho)\,\tau^{\hat{k}}(\varrho)}\,\breve{e}_{i},\\\\
 e_{\hat{i}}(\varrho)=-\frac{\pi^{\hat{0}}_{\hat{i}}(\varrho)}
{\pi(\varrho)\,\tau^{\hat{0}}(\varrho)-\pi^{\hat{0}}_{\hat{k}}(\varrho)\,\tau^{\hat{k}}(\varrho)}\,\breve{e}_{0}\,+
\pi^{-1}(\varrho)\left[\delta^{j}_{i}+
\frac{\tau^{j}(\varrho)\,\pi^{\hat{0}}_{\hat{i}}(\varrho)}
{\pi(\varrho)\,\tau^{\hat{0}}(\varrho)-\pi^{\hat{0}}_{\hat{k}}(\varrho)\,\tau^{\hat{k}}(\varrho)}\right]\,\breve{e}_{j}.
\end{array}
\label{W25}
\end{equation}
A {\em generalized transport} for deformed frame $e_{\hat{a}}$,
which includes both the Fermi-Walker transport and distortion of
$\underline{M}_{\,2}$, can be written in the form
\begin{equation}
\begin{array}{l}
\frac{de_{\phantom{a}\hat{a}}^{\,
\mu}}{ds}=\widetilde{\Phi}{}_{a}^{\phantom{a}b}\,e_{\phantom{a}\hat{b}}^{
\mu},
\end{array}
\label{WW10}
\end{equation}
where a {\em deformed acceleration tensor}
$\widetilde{\Phi}{}_{a}^{\phantom{a}b}$ concisely is given by
\begin{equation}
\begin{array}{l}
\widetilde{\Phi}=\frac{d\,\ln \pi}{ds}+\pi\,\Phi\,\pi^{-1}.
\end{array}
\label{WW11}
\end{equation}
Thus, we derive the tetrad fields
$e^{\phantom{a}\hat{s}}_{r}(\varrho)$~(\ref{W10}) and
$e_{\phantom{a}\hat{a}}^{\mu}(\varrho)$~(\ref{W25})  as a function
of {\em local rate} $\varrho$ of instantaneously change of a
constant velocity  (both magnitude and direction) of a massive
particle in $M_{4}$ under the unbalanced net force, describing
corresponding {\it fictitious graviton}. Then, the {\it fictitious
gravitino}, $\psi_{\hat{m}}^{\phantom{a}\alpha}(\varrho)$, will be
arisen under infinitesimal transformations of local
supersymmetry~(\ref{T25}), provided by the local parameters
$\zeta^{M}(a)$~(\ref{T6}).

\subsection{Involving the background semi-Riemann space $V_{4}$;
Justification for the introduction of the WPE} We can always choose
{\em natural coordinates} $X^{\alpha}(T,X,Y,Z)=(T,\,\vec{X})$ with
respect to the axes of the local free-fall coordinate frame
$S_{4}^{(l)}$ in an immediate neighbourhood of any space-time point
$(\breve{x}_{p})\in V_{4}$ in question of the background semi-
Riemann space, $V_{4}$, over a differential region taken small
enough so that we can neglect the spatial and temporal variations of
gravity for the range involved. The values of the metric tensor
$\breve{g}_{\mu\nu}$ and the affine connection
$\breve{\Gamma}^{\lambda}_{\mu\nu}$ at the point $(\breve{x}_{p})$
are necessarily sufficient information for determination of the
natural coordinates $X^{\alpha}(\breve{x}{}^{\mu})$ in the small
region of the neighbourhood of the selected point~\cite{W72}. Then
the whole scheme outlined in the previous subsections (a) and (b)
will be held in the frame $S_{4}^{(l)}$.  The  general {\em inertial
force} computed by~\cite{gago1} reads
\begin{equation}
\begin{array}{l}
\breve{\vec{f}}_{(in)}=
-\frac{m\vec{a}_{abs}}{\Omega^{2}(\overline{\varrho})\, \gamma_{q}}=
-\frac{\vec{e}_{f}}{\Omega^{2}(\overline{\varrho})\,
\gamma_{q}}|f^{\alpha}_{(l)}- m\frac{\partial X^{\alpha}}{\partial
\breve{x}{}^{\sigma}}\breve{\Gamma}{}^{\sigma}_{\mu\nu}
\frac{d\breve{x}{}^{\mu}}{dS} \frac{d\breve{x}^{\nu}}{dS}|.
\end{array}
\label{RLLL9}
\end{equation}
Whereas, as before, the two systems $S_{2}$ and $S_{4}^{(l)}$ can be
chosen in such a way as the axis $\vec{e}_{q}$ of $S_{(2)}$ lies
($\vec{e}_{q} =\vec{e}_{f}$) along the acting net force $\vec{f}=
\vec{f}_{(l)}+\vec{f}_{g(l)}$,  while the time coordinates in the
two systems are taken the same, $q^{0}=t=X^{0}=T.$ Here
$\vec{f}_{(l)}$ is the SR value of the unbalanced relativistic force
other than gravitational  and $\vec{f}_{g(l)}$ is the gravitational
force given in the frame $S_{4}^{(l)}$. Despite of totally different
and independent sources of gravitation and inertia, at
$f^{\alpha}_{(l)}=0$, the (\ref{RLLL9}) establishes the independence
of free-fall ($v_{q}=0$) trajectories of the mass, internal
composition and structure of bodies. This furnishes a justification
for the introduction of the WPE. A remarkable feature is that,
although the inertial force has a nature different than the
gravitational force, nevertheless both are due to a distortion of
the local inertial properties of, respectively,  2D
$\underline{M}_{\,2}$ and 4D-background space.

\subsection{The inertial effects in the background post Riemannian geometry}
If the nonmetricity tensor $N_{\lambda\mu\nu}=-{\cal
D}_{\lambda}\,g_{\mu\nu}\equiv -g_{\mu\nu\,; \lambda}$ does not
vanish, the general formula for the affine connection written in the
space-time components is~\cite{Popl}
\begin{equation}
\begin{array}{l}
\Gamma^{\rho}_{\phantom{a}\mu\,\nu}=\stackrel{\circ
}{\Gamma}{}^{\rho}_{\phantom{a}\mu\,\nu}
+K^{\rho}_{\phantom{k}\mu\nu}-N^{\rho}_{\phantom{k}\mu\nu}+\frac{1}{2}N_{(\mu\phantom{k}\nu)}^{\phantom{(i}\rho},
\end{array}
\label{RU1}
\end{equation}
where the metric alone determines the torsion-free  Levi-Civita
connection $\stackrel{\circ}{\Gamma}{}^{\rho}_{\phantom{a}\mu \nu}$,
$K^{\rho}_{\phantom{k}\mu\nu}:\phantom{a}=2Q_{(\mu\nu)}^{\phantom{(ij)}\rho}+
Q^{\rho}_{\phantom{k}\mu\nu}$ is the non-Riemann part - the affine
{\em contortion tensor}. The torsion, $Q^{\rho}_{\phantom{k}\mu\nu}=
\frac{1}{2}\,T^{\rho}_{\phantom{k}\mu\nu}=\Gamma^{\rho}_{\phantom{a}[\mu\,\nu]}$
given with respect to a holonomic frame, $d\,\vartheta^{\rho}=0$, is
a third-rank tensor, antisymmetric in the first two indices, with 24
independent components. We now compute the relativistic inertial
force for the motion of the matter, which is distributed over a
small region in the $U_{4}$ space and consists of points with the
coordinates $x^{\mu}$, forming an extended body whose motion in the
space, $U_{4}$, is represented by a world tube in space-time.
Suppose the motion of the body as a whole is represented by an
arbitrary timelike world line $\gamma$ inside the world tube, which
consists of points with the coordinates $\tilde{X}{}^{\mu}(\tau)$,
where $\tau$ is the proper time on $\gamma$. Define
\begin{equation}
\begin{array}{l}
\delta x{}^{\mu}=x{}^{\mu}-\tilde{X}{}^{\mu},\phantom{a}\delta
x^{0}=0,\phantom{a}u^{\mu}=\frac{d\,\tilde{X}{}^{\mu}}{d\,s}.
\end{array}
\label{I4}
\end{equation}
The {\em Papapetrou equation of motion for the modified momentum}
(\cite{Popl}-\cite{Mol2}) is
\begin{equation}
\begin{array}{l}
\frac{\stackrel{\circ }{\cal D}\,\Theta^{\nu}}{{\cal
D}\,s}=-\frac{1}{2}\,\stackrel{\circ
}{R}{}^{\nu}_{\phantom{j}\mu\sigma\rho}\,u{}^{\mu}\,
J^{\sigma\rho}-\frac{1}{2}\,N_{\mu\rho\lambda}\,K^{\mu\rho\lambda:\,\nu},
\end{array}
\label{I8}
\end{equation}
where $K^{\mu}_{\nu\lambda}$ is the contortion tensor,
\begin{equation}
\begin{array}{l}
\Theta^{\nu}=P^{\nu}+\frac{1}{u{}^{0}}\,\stackrel{\circ
}{\Gamma}{}^{\phantom{a}\nu}_{\mu\,\rho}\,(u{}^{\mu} \,J^{\rho
0}+N^{0 \mu\rho})-
\frac{1}{2u{}^{0}}\,K_{\mu\rho}^{\phantom{ik}\nu}\, N^{\mu\rho 0}
\end{array}
\label{I7}
\end{equation}
is referred to as the {\em modified 4-momentum},
$P^{\lambda}=\int\tau^{\lambda 0}\,d\,\Omega$ is the ordinary
4-momentum, $d\,\Omega:=d\,x^{4}$, and the following integrals are
defined:
\begin{equation}
\begin{array}{l}
 M^{\mu\rho}=u{}^{0}\,\int\tau^{\mu\rho}\,d\,\Omega, \quad
M^{\mu\nu\rho}=-u{}^{0}\,\int\delta
x{}^{\mu}\,\tau^{\nu\rho}\,d\,\Omega, \quad
N^{\mu\nu\rho}=u{}^{0}\,\int s^{\mu\nu\rho}\,d\,\Omega, \\
 J^{\mu\rho}=\int(\delta x{}^{\mu}\,\tau^{\rho 0}-\delta
x{}^{\rho}\,\tau^{\mu 0}+s^{\mu\rho
0})\,d\,\Omega=\frac{1}{u{}^{0}}(-M^{\mu\rho 0}+M^{\rho\mu
0}+N^{\mu\rho 0}),
\end{array}
\label{I6}
\end{equation}
where  $\tau^{\mu\rho}$ is the energy-momentum tensor for particles,
$s^{\mu\nu\rho}$ is the spin density. The quantity $J^{\mu\rho}$ is
equal to $\int(\delta x{}^{\mu}\,\tau^{kl}-\delta
x{}^{\rho}\,\tau^{\mu\lambda}+s^{\mu\rho\lambda})\,d\,S_{\lambda}$
taken for the volume hypersurface, so it is a tensor, which is
called the {\em total spin tensor}. The quantity $N^{\mu\nu\rho}$ is
also a tensor. The relation $\delta x{}^{0}=0$ gives
$M^{0\nu\rho}=0$. It was assumed that the dimensions of the body are
small, so integrals with two or more factors $\delta x{}^{\mu}$
multiplying $\tau^{\nu\rho}$ and integrals with one or more factors
$\delta x{}^{\mu}$ multiplying $s^{\nu\rho\lambda}$ can be
neglected. The {\em Papapetrou equations of motion for the spin}
(\cite{Popl}-\cite{Mol2} ) is
\begin{equation}
\begin{array}{l}
 \frac{\stackrel{\circ }{\cal D}}{{\cal
D}s}\,J^{\lambda\nu}=u^{\nu}\,\Theta^{\lambda}-u{}^{\lambda}\,\Theta^{\nu}+
K^{\lambda}_{\phantom{l}\mu\rho}\,N^{\nu\mu\rho}+\frac{1}{2}\,K_{\mu\rho}^{\phantom{ik}\lambda}\,
N^{\mu\nu\rho}-K^{\nu}_{\phantom{j}\mu\rho}\,N^{\lambda\mu\rho}-\frac{1}{2}\,K_{\mu\rho}^{\phantom{ik}\nu}\,N^{\mu\rho\lambda}.
\end{array}
\label{I5}
\end{equation}
Computing from~(\ref{I8}), in general, the relativistic inertial
force, exerted on the extended spinning body moving in the RC space
$U_{4}$, can be found to be
\begin{equation}
\begin{array}{l}
 \vec{f}_{(in)}(x)=
-\frac{m\vec{a}_{abs}(x)}{\Omega^{2}(\overline{\varrho})\,
\gamma_{q}}= -m\,\frac{\vec{e}_{f}}{\Omega^{2}(\overline{\varrho})\,
\gamma_{q}}\,\left|\frac{1}{m}\,f^{\alpha}_{(l)}- \frac{\partial
X^{\alpha}}{\partial\, x{}^{\mu}}\,\left[ \stackrel{\circ
}{\Gamma}{}^{\mu}_{\nu\lambda}\,u{}^{\nu}\,u^{\lambda}+
  \right.\right.\\\left.\left.
\frac{1}{u{}^{0}}\,\stackrel{\circ
}{\Gamma}{}^{\phantom{a}\mu}_{\nu\,\rho}\,(u{}^{\nu} \,J^{\rho
0}+N^{0
\nu\rho})-\frac{1}{2u{}^{0}}\,K_{\nu\rho}^{\phantom{ik}\mu}\,
N^{\nu\rho 0}+\frac{1}{2}\,\stackrel{\circ
}{R}{}^{\mu}_{\phantom{j}\nu\sigma\rho}\,u{}^{\nu}\,
J^{\sigma\rho}+\frac{1}{2}\,N_{\nu\rho\lambda}\,K^{\nu\rho\lambda:\,\mu}
 \right]\right|.
\end{array}
\label{RLLL10}
\end{equation}

\section{Concluding remarks}
We present a standard Lorentz code of motion in a new perspective of
supersymmetry. In this, we explore the intermediate, so-called, {\it
motion} state for a particle moving through the two infinitesimally
closed points of original space. The Schwinger transformation
function for these points is understood as the successive processes
of annihilation of a particle at initial point and time, i.e. the
transition from the initial state to the intermediate {\it motion}
state, and the creation of a particle at final point and time, i.e.
the subsequent transition from the intermediate {\it motion} state
to the final state. The latter is defined on the {\it master space},
MS \,$(\equiv \underline{M}_{\,2})$, which is prescribed to each
particle, without relation to every other particle. Exploring the
rigid double transformations of MS-SUSY, we derive SLC as the
individual code of a particle in terms of spinors referred to MS.
This allows to introduce the physical finite {\it time interval}
between two events, as integer number of the {\it duration time} of
atomic double transition of a particle from $M_{4}$ to
$\underline{M}_{\,2}$ and back. The theories with extended
$N_{max}=4$ supersymmetries, as renormalizable flat-space field
theories, if only such symmetries are fundamental to nature, lead to
the model of ELC in case of the apparent violations of SLC, the
possible manifestations of which arise in a similar way in all
particle sectors. We show that in the ELC-framework the propagation
of the superluminal particle could be consistent with causality, and
give a justification of forbiddance of Vavilov-Cherenkov
radiation/or analog processes in vacuum. In the framework of local
MS-SUSY, we address the inertial effects. The local MS-SUSY can only
be implemented if $\widetilde{\underline{M}}_{\,2}$ and
$\widetilde{M}_{4}$ are curved (deformed). Whereas the space
$\widetilde{M}_{4}$, in order to become on the same footing with the
distorted space $\underline{\widetilde{M}}_{\,2}$, refers to the
accelerated reference frame of a particle, without relation to other
matter fields. So, unlike gravitation,  a curvature of space-time
arises entirely due to the inertial properties of the
Lorentz-rotated frame of interest, i.e. a {\it fictitious
gravitation} which can be globally removed by appropriate coordinate
transformations.  The only source of graviton and gravitino,
therefore, is the acceleration of a particle, because the MS-SUSY is
so constructed as to make these two particles just as being the two
bosonic and fermionic states of a particle of interest in the
background spaces $M_{4}$ and $\underline{M}_{\,2}$, respectively,
or vice versa. Therefore, a coupling of supergravity with matter
superfields evidently is absent in resulting theory. Instead, we
argue that a deformation/(distortion of local internal properties)
of MS is the origin of inertia effects that can be observed by us.
In the framework of classical physics we briefly discuss the model
of inertia effects and go beyond the hypothesis of locality. This
allows to improve essentially the relevant geometrical structures
referred to the noninertial frame in Minkowski space-time for an
arbitrary velocities and characteristic acceleration lengths.
Despite the totally different and independent physical sources of
gravitation and inertia, this approach furnishes justification for
the introduction of the WPE. Consequently, we relate the inertia
effects to the more general post-Riemannian geometry.


\begin{thebibliography}{}

\bibitem{Drake}
S. Drake,  Galileo at work, Chicago, University of Chicago Press (1978),\\
(http://en.wikipedia.org/wiki/Galileo).

\bibitem{Newt}
I. Newton, Philosophiae Naturalis Principia Mathematica, (1687),\\
(http://plato.stanford.edu/entries/newton-principia).



\bibitem{Nort}
J. Norton, What was Einstein's principle of equivalence?, {\it Stud.
Hist. Phil. Sci.} {\bf 16} 203 (1985).


\bibitem{gago1}
G. Ter-Kazarian, Spacetime deformation induced inertia effects, {\it
Advances in Mathematical Physics}, Vol. 2012, Article ID 692030, 41
pages, doi:10.1155/2012/692030, Hindawi Publ. Corporation (2012).



\bibitem{gago2}
G. Ter-Kazarian, Two-step spacetime deformation induced dynamical
torsion, Class. Quantum Grav., {\bf 28}, 055003 (19pp.), (2011);
arXiv:1102.2491[gr-qc].


\bibitem{DHM}
H.K. Dreiner, H.E. Haber and S.P. Martin, Supersymmetry, CUP draft
Sept. pp.1-272 (2004).

\bibitem{FF}
P. Fayet and S. Ferrara, Supersymmetry, Physics Reports, {\bf 32}
pp.249-334 (1977).

\bibitem{Nw}
P. van Nieuwenhuizen, Supergravity, Physics Reports, {\bf 68}
pp.189-398 (1981).


\bibitem{WB}
J. Wess and J. Bagger, Supersymmetry and Supergravity,  Princeton
University Press, Princeton, New Jersey, pp.1-181 (1983).

\bibitem{GGRS}
M S.J. Gates, M.T. Grisaru, M. Ro\u{c}ek and W. Siegel, Superspace:
Or One Thousand and One Lessons in Supersymmetry, Benjamin/Cumming,
London, pp.1-548 (1983).


\bibitem{Nil}
H.P. Nilles, Supersymmetry, Supergravity and Particle Physics,
Physics Reports, {\bf 110}  pp.1-162 (1984).

\bibitem{Soh}
M.F. Sohnius, Introducing Supersymmetry, Physics Reports, {\bf 128}
pp.39-204 (1985).


\bibitem{We}
P. West, Introduction to Supersymmetry and Supergravity, World
Scientific, Singapure, pp.1-289 (1987).

\bibitem{Ja}
M. Jacob, Supersymmetry and Supergravity, North Holland, Amsterdam,
(1987).


\bibitem{BT}
H. Baer and X, Tata,  Weak Scale Supersymmetry: From Superfields to
Scattering Events, (Cambridge: Cambridge University Press) (2006).

\bibitem{Ai}
I.J.R. Aitchison,  Supersymmetry in particle physics: an elementary
introduction, SLAC Report SLAC-R-865 (2007).

\bibitem{Bi}
P. Binetruy, G. Girardi, R. Rrimm, Supergravity couplings: a
geometric formulation, Physics Reports, {\bf 343}  pp.255-462
(2001).

\bibitem{Dr}
M. Drees,  Supersymmetry, (Oxford: Oxford University Press) (2007).

\bibitem{Din}
M. Dine,  Supersymmetry and String Theory: Beyond the Standard
Model, (Cambridge: Cambridge University Press) (2007).

\bibitem{NFF}
P. van Nieuwenhuizen, D.Z. Freedman and S. Ferrara, Progress toward
a Theory of Supergravity, Phys. Rev., {\bf D13}  3214-3218 (1976).

\bibitem{DZ}
S. Deser and B. Zumino, Consistent Supergravity, Phys. Lett. {\bf
B62}  335 (1976).


\bibitem{Moh}
R.N. Mohapatra, Unification and Supersymmetry, The Frontiers of
Quark-Lepton Physics, 3th ed., Springer-Verlag New York, Inc.,
pp.1-421 (2002).

\bibitem{DM}
M. Dine and J.D. Mason, Supersymmetry and its dynamical breaking,
Rep. Prog. Phys., {\bf 74}, pp.1-29  056201 (2011).

\bibitem{WZ}
J. Wess and B. Zumino, A Lagrangian model invariant under supergauge
transformations, Phys. Lett. {\bf 49B}  52 (1974).


\bibitem{W1}
S. Weinberg,  Nonlinear Realizations of Chiral Symmetry, Phys. Rev.,
{\bf 166} 1568 (1968).

\bibitem{CWZ}
S. Coleman, J. Wess and B. Zumino, Structure of phenomenological
Lagrangians.I, Phys. Rev. {\bf 177} 2239 (1969).



\bibitem{CCWZ}
C. Callan, S. Coleman, J. Wess and B. Zumino,  Structure of
phenomenological Lagrangians. 2, Phys. Rev. {\bf 177} 2247 (1969).


\bibitem{V}
D.V. Volkov,  Phenomenological Lagrangians, Sov. J. Particles and
Nuclei {\bf 4} 1-17 (1973).

\bibitem{O}
V.I. Ogievetsky, in Proc. of X-th Winter School of Theoretical
Physics in Karpacz, Wroclaw (1974).

\bibitem{SS}
A. Salam and J. Strathdee, Supergauge Transformations, Nucl. Phys.
{\bf B76} 477 (1974).


\bibitem{La}
L.D.Landau and E.M. Lifshitz, Electrodynamics of Continuous Media,
Theoretical physics, vol.VIII. Moscow, Nauka (1992).


\bibitem{Syn}
J.L. Synge,  Relativity: The General Theory, North-Holland,
Amsterdam (1960).

\bibitem{MTW}
C.W. Misner, K.S. Thorne  and J.A. Wheeler,  Gravitation (Freeman,
San Francisco) (1973).


\bibitem{Hehl7}
F.W. Hehl, W.-T. Ni, Inertial effects of a Dirac particle, Phys.
Rev. D, {\bf 42} pp.2045-2048 (1990).

\bibitem{HL} F.W. Hehl, J. Lemke and E.W. Mielke,  Two
    lectures on fermions and gravity, In: {\it Geometry and Theoretical
  Physics}, Debrus J and Hirshfeld A C (eds.) (Springer, Berlin) p.
  56 (1991).


\bibitem{MB1}
B. Mashhoon, Length measurement in accelerated systems, Ann. der
Physik, {\bf 514} 532 (2002).

\bibitem{MB2}
B. Mashhoon,  Necessity of acceleration-induced nonlocality, Ann.
der Physik, {\bf 523}  226 (2011);\,  ibidem  {\bf 520} 705 (2008);
arXiv:hep-th/0507157; ibidem {\bf 12} 586 (2003);
arXive:0309124[hep-th]; ibidem {\bf 198} 9 (1995).


\bibitem{MB5}
B. Mashhoon, Limitations of spacetime measurements, Phys. Lett. A,
{\bf 143}, pp.176-182 (1990).



\bibitem{MB6}
B. Mashhoon, The hypothesis of locality in relativistic physics,
Phys. Lett. A, {\bf 145} pp.147-153 (1990).


\bibitem{MB7}
B. Mashhoon, Neutron Interferometry in a Rotating Frame of
Reference, Phys. Rev. Lett., {\bf 61} pp.2639-2642 (1988).

\bibitem{MB8}
B. Mashhoon, On the coupling of intrinsic spin with the rotation of
the earth, Phys. Rev. Lett., {\bf 198} pp.9-13 (1995).

\bibitem{MFU}
J.W. Maluf, F.F. Faria   and S.C. Ulhoa,  On reference frames in
spacetime and gravitational energy in freely falling frames, Class.
Quantum Grav., {\bf 24}  pp.2743-2753 (2007);
arXiv:0704.0986[gr-qc].

\bibitem{MF}
J.W. Maluf and F.F. Faria,  On the construction of Fermi-Walker
transported frames, Ann. der Physik, {\bf 520} pp.326-335 (2008);
arXiv:0804.2502[gr-qc].


\bibitem{Marz}
K.-P. Marzlin,  What is the reference frame of an accelerated, Phys.
Lett. A,  {\bf 215} pp. 1-6 (1996).

\bibitem{Ni77}
W.-T. Ni,  On the Proper Reference Frame and Local Coordinates of an
Accelerated Observer in Special Relativity, Chinese J. Phys., {\bf
15} pp.51-55 (1977).

\bibitem{Ni78}
W.-Q. Li  and W.-T. Ni, On an Accelerated Observer with Rotating
Tetrad ' in Special Relativity, Chinese J. Phys. {\bf 16} 214
(1978).

\bibitem{Ni78D}
W.-T. Ni  and M. Zimmermann,  Inertial and gravitational effects in
the proper reference frame of an accelerated, rotating observer,
Phys. Rev. D, {\bf 17} 1473 (1978).

\bibitem{Gron}
F. Gronwald and F.W. Hehl, On the Gauge Aspects of Gravity, Proc. of
the 14th Course of the School of Cosmology and Gravitation on
Quantum Gravity, held at Erice, Italy, Eds. Bergmann P G, de Sabbata
V, and Treder H.-J., World Scientific, Singapore (1996);
arXiv:9602013[gr-qc].


\bibitem{W72}
S. Weinberg, Gravitation and Cosmology, J. W. and Sons, New York
(1972).


\bibitem{Popl}
N.J. Poplawski,  Spacetime and fields, pp.1-114 (2009);
[gr-qc/0911.0334].

\bibitem{Pa}
A. Papapetrou,   Einstein's Theory of Gravitation and Flat Space,
Proc. Roy. Irish Acad. A, {\bf 52} pp.11-23 (1948);
 Proc.R.Soc. A, {\bf 202} 248 (1951);  Lectures on
general relativity, (Reidel D)(ISBN 9027705402) (1974).


\bibitem{Be}
P.G. Bergmann   and R. Thompson,  Spin and angular momentum in
general relativity, Phys. Rev. {\bf 89} 400 (1953).

\bibitem{Mol2}
C. M{\o}ller,   Ann. Phys. (NY), On the Localization of the Energy
of a Physical System in the General Theory of Relativity, {\bf 4}
pp.347-371 (1958).


\end{thebibliography}
\end{document}